\documentclass[12pt]{iopart}

\usepackage{bm, graphicx,color, slashed}
\usepackage{amssymb, bbm}
\usepackage[all]{xy}
\xyoption{import}

\def\<{\langle}
\def\>{\rangle}

\def \I{\mathbbm{1}}
\def\H{ {\mathcal H} }
\def\C{ {\mathcal C} }
\def\D{ {\mathcal D} }

\def\T{ {\mathcal T} }
\def\P{ {\mathcal P} }
\def\A{ {\mathcal A} }
\def\B{ {\mathcal B} }

\def\sys{\widehat{\psi}_\A}
\begin{document}
\bibliographystyle{amsplain}
\title[Continuum tensor networks and symmetries]{Continuum tensor network field states, path integral representations and the encoding of spatial symmetries.}

\author{David Jennings}
\address{Controlled Quantum Dynamics Theory Group, Level 12, EEE, Imperial College London, London SW7 2AZ, United Kingdom}
\author{Christoph Brockt}
\address{Leibniz Universit\"at Hannover, Institute of Theoretical Physics, Appelstra{\ss}e 2, D-30167 Hannover, Germany}
\author{Jutho Haegeman} 
\address{Vienna Center for Quantum Science and Technology, University of Vienna, Austria}
\author{Tobias J.\ Osborne} 
\address{Leibniz Universit\"at Hannover, Institute of Theoretical Physics, Appelstra{\ss}e 2, D-30167 Hannover, Germany}
\author{Frank Verstraete}
\address{Vienna Center for Quantum Science and Technology, University of Vienna, Austria}

%\ead{david.b.jennings@gmail.com}

\begin{abstract}
A natural way to generalise tensor network variational classes to quantum field systems is via a continuous tensor contraction. This approach is first illustrated for the class of quantum field states known as continuous matrix-product states (cMPS). As a simple example of the path-integral representation we show that the state of a dynamically evolving quantum field admits a natural representation as a cMPS. A completeness argument is also provided that shows that all states in Fock space admit a cMPS representation when the number of variational parameters tends to infinity. Beyond this, we obtain a well-behaved field limit of projected entangled pair states (PEPS) in two dimensions that provide an abstract class of quantum field states with natural symmetries. We demonstrate how symmetries of the physical field state are encoded within the dynamics of an auxiliary field system of one dimension less. In particular, the imposition of Euclidean symmetries on the physical system requires that the auxiliary system involved in the class' definition must be Lorentz-invariant. The physical field states automatically inherit entropy area laws from the PEPS class, and are fully described by the dissipative dynamics of a lower dimensional virtual field system. Our results lie at the intersection many-body physics, quantum field theory and quantum information theory, and facilitate future exchanges of ideas and insights between these disciplines.
\end{abstract}

%\noindent{\it Keywords}: entanglement, quantum information, many-body physics, quantum fields

\maketitle

\section{Introduction}

The quantum states that we observe in nature are highly atypical as compared to a state randomly chosen from the full configuration Hilbert space $\H$ \cite{poulin:2011a}. Indeed, observable states comprise only a tiny submanifold of $\H$ --- the \textit{physical corner of Hilbert space} --- whose points exhibit highly nongeneric features such as nontrivial clustering of correlations and entropy areas laws \cite{hastings:2005b, eisert:2010b}. It is extremely desirable to develop an efficient  parametrisation of this manifold as this would considerably ameliorate the computational costs of solving physical models and provide new analytical tools for the study of quantum field systems. Indeed, even a partial parametrisation of the physical corner provides a powerful tool as it supplies a variational class useful for the description of low-energy physics.

The canonical example of such a class of quantum states appears in the setting of one-dimensional lattices. There the class of \emph{matrix product states} (MPS) \cite{fannes:1992a} has enjoyed remarkable success, not simply for the calculation of physical properties of strongly interacting lattices, but also for such things as the classification of quantum phases, providing a natural foliation of states in terms of entanglement, and the construction of exactly solvable models \cite{verstraete:2008review, wolf:2005a,verstraete:2005a}. It is also well-established that MPS satisfy two important criteria. Firstly they constitute a \textit{complete} class of quantum states, in the sense that by increasing a ``bond dimension'' $D$ one can capture any pure quantum state of the system. Secondly the class is \textit{efficient} in the sense that the computational cost of calculating expectation values scales polynomially in the number of variational parameters. 

The MPS class has provided a fruitful basis for generalisations: by understanding the structure of quantum entanglement in such states they have inspired several powerful extensions to higher dimensions and different geometries. The two most prominent examples for higher-dimensional lattice systems are the \emph{projected entangled-pair states} (PEPS) \cite{verstraete:2004a}  and the \emph{multiscale entanglement renormalisation ansatz} (MERA) \cite{vidal:2006a, vidal:2007a}. Both of these variational classes have proved invaluable in the investigation of strongly correlated physics. So far, however, all of these results have been restricted to the lattice setting; the study of continuous quantum systems using these classes (more generally referred to as \emph{tensor networks}) has traditionally proceeded by first discretising the system on the lattice and then employing them as a variational ansatz. 

Continuum systems bring considerable difficulties when it comes to variational computations because optimisations can be dominated by UV physics at the expense of infra-red physics which ruins the estimation of observables of physical interest \cite{feynman:1987a}. Remarkably, both of these difficulties have been overcome with the introduction of special continuum versions of the MPS and MERA classes \cite{verstraete:2010a, osborne:2010a, haegeman:2010a, haegeman:2011a}. The cMPS class is remarkable in that it requires (in the translation-invariant case) only a finite number of variational parameters to specify, but is expected, by analogy with the discrete case, to be both efficient and complete in the sense already described. Further studies have also established that cMPS and cMERA are not disrupted by the presence of UV divergences \cite{haegeman:2010a, haegeman:2011a}. Here we argue that the most natural systematic way to achieve this is to replace the tensor contraction with a path integral over some now continuous auxiliary degrees of freedom. 

We should emphasize that the goal of this work is not a new formulation of quantum fields, but instead the construction of a manifold of quantum field states that possess natural properties. In particular, we wish to extend results obtained in one spatial dimensions to higher dimensions, and develop a novel toolkit (analytical and computational) for the study of strongly interacting, and highly correlated two and three dimensional quantum systems.

In what follows, we construct a field limit of both 1-d and 2-d tensor networks, show how tensor-contractions naturally pass over into path integral over virtual degrees of freedom, and then develop the field limit of a lattice PEPS. This generalisation takes the same functional form as the one-dimensional cMPS and manifestly exhibits spatial (e.g. rotational) symmetries. The derivation via a sequence of lattice PEPS means that the resultant class of field states automatically obey entropy area laws. Furthermore, the imposition of spatial symmetries on the physical field state is obtained by encoding the symmetry into the dynamics of an auxiliary boundary system with the novel result that the dynamics of the boundary system is given by the imaginary time evolution of a Lorentz invariant system of one lower spatial dimension.

\section{Background: matrix product states, tensor networks, and coherent state path integrals}\label{sec:background}
Here we review the MPS class and sketch some of its properties. Our intent is to make this paper accessible to those with a diversity of backgrounds, so we provide all the necessary prerequisite material and references needed to follow our argument here. Readers with a familiarity with MPS and the DMRG are invited to skim this section lightly to fix notation.

We begin by recalling that any bipartite pure quantum state $|\psi_{\A \B}\>$ admits a Schmidt decomposition $|\psi_{\A \B}\> = \sum_k \sqrt{\lambda_k} |k\>_\A \otimes |k\>_\B$ for some set of local bases of $\A$ and $\B$. For pure states  $|\psi\>$  of one-dimensional quantum spin systems $\A_1 \A_2 \cdots \A_n$ with local dimension $d$, we may perform a Schmidt decomposition iteratively on the bipartitions $[\A_1, \A_1']$, $[(\A_1\A_2), (\A_1\A_2)'], \cdots$ $[\A_n', \A_n]$  (where $X'$ is the complement of $X$ and $[X, Y]$ denotes the particular bipartite split) to obtain the MPS representation \cite{vidal:2003b}
\begin{equation}\label{eq:mps}
	|\psi\rangle = \sum_{j_1, \ldots, j_n = 0}^{d-1} \langle \omega_L|A^{j_1}A^{j_2} \cdots A^{j_n}|\omega_R\rangle |j_1j_2\cdots j_n\rangle.
\end{equation}
Here $A^{j_k}$, $j_k = 0, 1, \ldots, d-1$, is a collection of $d$ matrices of size $D_{k-1}\times D_{k}$, $\langle\omega_L|$ is a row vector of dimension $D_0$, and $|\omega_R\rangle$ is a column vector of dimension $D_n$. The dimensions $D_k$ are called the \emph{bond dimensions} of the MPS and characterize the degree of entanglement entropy across a cut at site $k$. This construction shows that MPS are an\ \emph{expressive} class, meaning that \emph{any} state may be represented as an MPS with a sufficiently large choice of the $D_k$s (the argument applies to \emph{any} pure state). However, in most implementations we simply assume that the bond dimension is constant and truncate it at some value $D_k=D$, which acts as a refinement parameter for this class.

Matrix product state representations (\ref{eq:mps}) possess several remarkable properties. The first, and most important, is that they provide an efficient parametrisation of naturally occurring states \cite{hastings:2007a, osborne:2005d, schollwoeck:2005a, schollwock:2011a}; MPS very efficiently represent both the ground states of models with a spectral gap and also the non-equilibrium dynamics of any quantum spin chain. The second property is that they possess an entropy area law \cite{eisert:2010b}, meaning that the von Neumann entropy of any contiguous block of spins is bounded above by a constant, i.e., the size of the boundary. Another important property of MPS is a \emph{gauge} degree of freedom, so  they supply an over-complete parametrisation. 

A matrix product state (\ref{eq:mps}) is an example of a quantum state known as a \emph{tensor network state} (TNS). To define a TNS one first associates a finite graph $G = (V, E)$ with the quantum system where the physical degrees of freedom, which are of dimension $d$, live on the vertices $V$, and the edges $E$ encode \emph{auxiliary} degrees of freedom. To each vertex $v$ we associate a tensor $A_{\alpha_{e_1}\alpha_{e_2}\cdots \alpha_{e_{d_v}}}^{j_v}$ with $d_v+1$ indices, where $d_v$ is the \emph{degree} of the vertex $v$. Each index $\alpha_{e}$ is associated with a corresponding edge $e \in E$ incident with the vertex $v$ and takes values $1, 2, \ldots, D_{e}$; these are the \emph{auxiliary bond indices}. The index $j_v$ is the \emph{physical index} and takes values $0, 1, \ldots, d-1$. The TNS corresponding to this arrangement of tensors is then given by
\begin{equation}
	|\psi\rangle = \sum_{j_{v_1},j_{v_2},\cdots, j_{v_{|V|}}}\mathcal{C}( A^{j_{v_1}} A^{j_{v_2}} \cdots A^{j_{v_{|V|}}})|j_{v_1} j_{v_2} \cdots j_{v_{|V|}}\rangle,
\end{equation}  
where $\mathcal{C}$ denotes the contraction of all the auxiliary indices, i.e., each pair of tensor indices associated with each edge are separately summed. Such TNSs may be represented pictorially where we draw a ``leg'' for each index of each tensor and join contracted indices with lines. Physical indices are drawn as unpaired arrows. For example, the simple tensor network resulting from the multiplication of two matrices $\sum_{\beta} A_{\alpha \beta}B_{\beta \gamma}$, is represented by:
\begin{equation*}
	\hspace{2cm}\xymatrix{
	 &*+[F-:<3pt>]\txt{$A$}\ar[l]_{\alpha}\ar@{-}[r]^{\beta} &*+[F-:<3pt>]\txt{${B}$}\ar[r]^{\gamma}&.	}
\end{equation*}

In the case of an MPS we associate with each tuple of matrices $A^{j_k}$, regarded as a \emph{three-index tensor} $[A^{j_k}]_{\alpha_{k-1}\alpha_{k}}$, the diagram according to
\begin{equation*}
	\xymatrix{
	 & & &\\
	  [A^{j_k}]_{\alpha_{k-1}\alpha_{k}}&\equiv&&*+[F-:<3pt>]\txt{$A$}\ar[r]^{\alpha_{k}}\ar[u]_{j_k}\ar[l]_{\alpha_{k-1}}&, \\
	}
\end{equation*}
In the pictorial representation the coefficient of $|j_1j_2\cdots j_n\rangle$ for an MPS is depicted as
\begin{equation*}
\hspace{-3cm}	\xymatrix{
	 & & & & & & \\
	 \langle\omega_L|A^{j_1}A^{j_2} \cdots A^{j_n}|\omega_R\rangle& = &*+[o][F-]{\omega_L}\ar@{-}[r] &*+[F-:<3pt>]\txt{$A$}\ar@{-}[r]^{\alpha_{1}}\ar[u]_{j_1} &*+[F-:<3pt>]\txt{$A$}\ar[r]^{\alpha_{2}}\ar[u]_{j_2}& \cdots & *+[F-:<3pt>]\txt{$A$}\ar[l]_{\alpha_{n-1}}\ar[u]_{j_n}&*+[o][F-]{\omega_R}\ar@{-}[l]}
\end{equation*}

The contraction involved in the definition of a tensor network state may also be expressed in terms of a path integral. To do this we define the following discrete ``action''
\begin{equation}
	S [ (\alpha_1, \alpha_2 \ldots, \alpha_{|E|}); (j_1, \ldots, j_{|V|}) ] \equiv \sum_{v\in V} -i\log (A^{j_v}_{\alpha_{e_1}\alpha_{e_2}\cdots \alpha_{e_{d_v}}}),
\end{equation}
With this definition, the TNS is given by 
\begin{equation}\label{eq:dfpi}
	|\psi\rangle = \int \mathcal{D}\boldsymbol{\alpha}\,\mathcal{D}\boldsymbol{j} \, e^{iS[\boldsymbol{\alpha}, \boldsymbol{j}]} |\boldsymbol{j}  \rangle,
\end{equation}
where $\int \mathcal{D}\boldsymbol{\alpha}\mathcal{D}\boldsymbol{j}$ anticipates the continuum, and denotes here a discrete \emph{sum over all paths} $\boldsymbol{\alpha} = (\alpha_1, \alpha_2, \ldots, \alpha_{|E|})$ and $(j_1, \ldots, j_{|V|})$ with $\alpha_k \in \{1, \ldots, D\}$ and $j_k \in \{1, 2, \ldots, d\}$. We note that this perspective on the discrete MPS also finds connections with other discrete path integral representations for unitary operators coming from measurement-based quantum computation \cite{Beaudrap:2008aa} (moreover it would also be of interest to see if traditional perturbative techniques of path integrals could find application in wholly discrete contexts).

In the next section we are faced with taking the continuum limit of these discrete structures. Intuitively speaking, the way in which we obtain the continuum limit of a TNS is to choose the tensors $A^j$ so that as the spacing between the sites goes to zero the density of particles/excitations in the system remains constant. More concretely, let us first imagine a classical setting, and a state $|00 \cdots 0110 \cdots 0\>$ that encodes the location of particles along a discrete one-dimensional system in terms of a length-$N$ string ``$00\cdots 0110 \cdots 0$''. Here ``$1$'' can be viewed as denoting that a single particle is present at a single location, while ``$0$'' represents that no particle is present. Moreover, one can imagine that the presence or absence of a particle occurs with some fixed probability so that any particular string occurs (classically) with a binomial distribution. By coarse-graining this string into cells of finite length we can count the number of particles $n(x)$ in a particular cell $x$ and therefore define an expected particle density $\<n(x)\>$ for the cell. Crucially, the passage to the continuum involves taking $N\rightarrow \infty$ while simultaneously sending $p\rightarrow 0$ at the same rate so as to keep $\<n(x)\>$ finite.

Essentially the same idea is employed for the fully quantum non-commuting case, with the demand that particles appear in the state  at such a rate (quantified by the label $j=0,1,2\dots$ on the MPS operators $\{A^j\}$) as to ensure finite expectation value for local Hermitian observables in the continuum limit.
For the one-dimensional case this is achieved \cite{verstraete:2010a, osborne:2010a} by choosing
\begin{eqnarray}\label{discrete}
		A^0 &= \mathbb{I} + \epsilon Q \nonumber \\
		A^1 &= \epsilon R \nonumber\\
		A^2 & = \frac{(\epsilon R)^2}{\sqrt{2!}} \nonumber \\
		&\vdots  \nonumber \\
		A^k &= \frac{(\epsilon R)^k}{\sqrt{k!}} \nonumber \\
		&\vdots  
\end{eqnarray}
where $Q$ and $R$ are arbitrary $D\times D$ matrices and $\epsilon$ is the lattice spacing. In particular $A^0(x)$ should be interpreted as ``no particle created at $x$'', while $A^1(x)$ should be interpreted as ``a single particle created at $x$''. We'll see in the next section that with this choice of $A^j$s the path integral (\ref{eq:dfpi}) reduces in the limit $\epsilon \rightarrow 0$ to a standard path integral, and the particular scaling in $\epsilon$ ensures finite expectation values of local observables. We shall show that a similar recipe works for any sufficiently regular lattice.

\section{Path integrals and continuous matrix product states}\label{sec:picmps}
Continuous matrix product states are a variational class of states for one-dimensional quantum fields. We focus on the bosonic case with field annihilation and creation operators $\sys(x)$ and $\sys^\dagger(x)$ such that $[\sys(x), \sys^\dagger(y)]=\delta(x-y)$. A cMPS is then defined in terms of the quantum field $\mathcal{A}$ and an auxiliary $D$-level quantum system $\mathcal{B}$ by
\begin{equation}\label{eq:cmps2}
\hspace{-2.5cm}|\chi \> = \langle \omega_L| \P \exp\left[-i\int_0^l K(s)\otimes\mathbb{I} + iR(s)\otimes \sys^\dag(s) - iR^\dag(s)\otimes \sys (s) \,ds\right]|\omega_R\rangle |\Omega_{\mathcal{A}}\rangle,
\end{equation}
where $K$ is a $D\times D$ hermitian matrix and $R$ is $D\times D$ complex matrix, $|\omega_{L,R}\>$ are $D$-dimensional states of the auxiliary system $\B$, $\sys (s)$ is a bosonic field operator on the physical system $\A$, $|\Omega_{\mathcal{A}}\rangle$ is the Fock vacuum, and $\mathcal{P}$ denotes path ordering. 

The above form (\ref{eq:cmps2}) can be derived directly from the discrete MPS data provided in (\ref{discrete}) and constructing the MPS state on $N$ sites as in (\ref{eq:mps}). One then makes use of the particular form of matrices, together with the product expansion formula for time-ordered exponentials
\begin{equation}
\P \exp \left [\int_a^b ds F(s) \right ] = \lim_{\epsilon \rightarrow 0} [e^{\epsilon F(s_N)} e^{\epsilon F(s_{N-1})} \cdots e^{\epsilon F(s_1)}],
\end{equation}
with $s_N = b$ and $s_1 =a$ and $\epsilon = L/N$ for some fixed length $L$. The continuum contributions can be extracted by expanding exponentials and grouping terms that are linear in $\epsilon$. Finally the $\epsilon \rightarrow 0$ limit yields expression (\ref{eq:cmps2}) with the relation between $Q$ and $K$ given by $iK = Q + \frac{1}{2} R^\dagger R$.

\subsection{A path integral for the auxiliary system}
We can reformulate the cMPS state (\ref{eq:cmps2}) so that expectation values for the auxiliary system are recast as path-integral expressions, using standard techniques. The motivation for this is two-fold: firstly, to facilitate the passage to higher-dimensional cMPS states; and secondly, to make manifest the symmetries of the physical state in terms of symmetries of an action for the auxiliary system. Our discussion is centred on the case of a single bosonic field in (1+1) dimensions; the generalisation to spinor and vector fields follows easily, and we only comment on the modifications required.

Write a basis for the Hilbert space $\mathcal{H}_{\mathcal{B}}$ of $\mathcal{B}$ as $\{|j\rangle\,|\, j = 0, 1, \ldots, D-1 \}$. We enlarge this space via second quantisation, and introduce bosonic annihilation and creation operators $b_j$ and $b_j^\dag$ according to the canonical commutation relations $[b_j, b_k^\dag] = \delta_{j,k}, \quad j = 0, 1, \ldots, D-1$, with all other commutators vanishing, or fermionic annihilation and creation operators $c_j$ and $c_j^\dag$ according to the canonical anticommutation relations $	\{c_j, c_k^\dag\} = \delta_{j,k}, \quad j = 0, 1, \ldots, D-1$, with all other anticommutators vanishing. The Hilbert space for our enlarged auxiliary system is that of the Fock space $\mathfrak{F}_{\pm}(\mathcal{H}_{\mathcal{B}})$, where the $\pm$ subscript indicates either bosonic or fermionic Fock space.

The connection between $\mathcal{H}_{\mathcal{B}}$ and our enlarged Fock space $\mathfrak{F}_{\pm}(\mathcal{H}_{\mathcal{B}})$ is made, in the bosonic case, by identifying $\mathcal{H}_{\mathcal{B}}$ with the single-particle sector via $|j\rangle_\mathcal{B} \equiv b_j^\dag |\Omega\rangle_{\mathcal{B}}$,
or, in the fermionic case,	$|j\rangle_\mathcal{B} \equiv c_j^\dag |\Omega\rangle_{\mathcal{B}}$,
where $|\Omega\rangle_{\mathcal{B}}$ is the Fock vacuum. We identify, whenever clear from the context, states $|\omega\rangle \in \mathcal{H}_{\mathcal{B}}$ with their single-particle counterparts in $\mathfrak{F}_{\pm}(\mathcal{H}_{\mathcal{B}})$.

Using this embedding, a cMPS (\ref{eq:cmps2}) is equivalent, in the bosonic case, to
\begin{equation}
\hspace{-1.5cm}|\chi\> = \<\omega_L | U(l,0)|\omega_R\> |\Omega_\A\> = \<\omega_L | \P \exp \left[-i\int_0^l F(s)\, ds \right]|\omega_R\> |\Omega_\A\>,
\end{equation}
where $F$ is a one-parameter set of field operators on $\A \B$, generated by $U(l,0)$ and given is by
\begin{eqnarray*}
\hspace{-2cm}F(s) =  \sum_{j,k=1}^D K^{jk}(s) b_j^{\dag} b_k  \otimes\I + iR^{jk}(s)  b_j^{\dag}b_k\otimes \sys^\dag(s) -iR^{*\, kj}(s)  b_j^{\dag}b_k\otimes \sys(s),
\end{eqnarray*}	
This equivalence of definitions follows from the fact that $F(s)$ is particle-number conserving on system $\B$ (i.e. its action on $\B$ is through terms of the form $b_j^\dagger b_k$ only), and so we remain in the single-particle sector throughout. The fermionic version is identical except that $b_j$ operators are replaced with $c_j$s. 

The parameter $s$ can be regarded as a \emph{time coordinate} for the auxiliary system. We then obtain a path integral by dividing $[0,l]$ into small intervals $s_0=0, s_1, s_2, \dots s_N=l$ with uniform spacing $s_{k+1}-s_k = \epsilon$, so that $U(l,0) = U(l,l-\epsilon)U(l-\epsilon,l-2\epsilon)\cdots U(\epsilon,0)$, and then inserting resolutions of the identity between each term. Our choice of resolution is, in the bosonic case, in terms of coherent states of the auxiliary system, defined as $|\phi_k\> = \exp[ \phi_k b_k^\dagger - \phi^*_k b_k]|\Omega_\B\>$:
\begin{equation}\label{eq:boscohres}
\I = \frac{1}{\pi^N} \int \prod_k d^2\phi_k \,|{}\otimes_k \phi_k\rangle\<\otimes_k \phi_k|,
\end{equation}
where $N = l/\epsilon$.
In the fermionic case we exploit fermion coherent states of the form $|\phi_k\> = \exp[ c_k^\dagger\phi_k  - \phi^*_kc_k]|\Omega_\B\>$, where $\phi_k$ are now Grassmann-valued. Apart from the use of anticommuting Grassmann numbers the fermionic calculation mirrors the bosonic case in essentially all other respects; we thus focus on the details of the bosonic calculation and write out the fermionic case at the end.

After the resolution (\ref{eq:boscohres}) has been placed between each term we end up with a product of transition amplitudes of the form $\<\otimes_k \phi_k (s+\epsilon)|U(s+\epsilon ,s)|\otimes_{k'} \phi_{k'} (s)\> \approx \<\otimes_k \phi_k(s+\epsilon)| \I -i\epsilon F(s)|\otimes_{k'}\phi_{k'} (s)\>$. We then make use of the expression
\begin{equation}\label{coherentoverlap}
\hspace{-2.5cm}\<\otimes_k \phi_k(s+\epsilon)|\otimes_{k'} \phi_{k'} (s)\> = \exp \left[-\frac{1}{2} \sum_{k=1}^D  |\phi_k(s+\epsilon)|^2 + |\phi(s)|^2 - 2 \phi_k^*(s+\epsilon)\phi_k(s) \right]
\end{equation}
and the assumption that only smooth variations of $\phi_k(s)$ contribute, which allows us to expand the terms in the exponential and obtain, in the continuum limit $\epsilon \rightarrow 0$,
\begin{eqnarray}
|\chi\> &=& \int \prod_k \D^2 \phi_k\, \P \exp \left[ i \widehat{S} (\phi_k, \phi_k^*) \right] |\Omega_\A\>,
\end{eqnarray}
where the path integral is over $D$ complex fields and $\widehat{S}$ is an operator-valued action given by
\begin{equation}
	\widehat{S} = \int ds \left(i \phi^\dagger \partial_s \phi - \phi^\dagger K\phi - i(\phi^\dagger R\phi)\sys^\dagger + i(\phi^\dagger R^\dag\phi)\sys\right),
\end{equation}  
where we abbreviate $\{ \phi_k\}$ as a vector $\phi$. However, since the field operator $\sys^\dagger(s)$ commutes with $\sys(s')$ and $\sys^\dag(s')$ at all other points $s'$ the ordering over the auxiliary time variable is trivial and we can simply write the path integral as
\begin{equation}\label{pathcmps}
|\chi\> = \int \D^2\phi\, \exp \left[ iS (\phi, \phi^\dagger) \right] |\Phi\>
\end{equation}
where $|\Phi\rangle$ is a \emph{physical} field coherent state 
\begin{equation}
	|\Phi\> \equiv \exp \left[\int \Phi(s) \sys^\dagger(s) - \Phi^*(s) \sys (s) \, ds \right] |\Omega_\A\>,
\end{equation}
$\Phi(s) = \phi^\dagger R\phi$, and the complex action $S$ is given by
\begin{equation}\label{cmpsaction}
	S(\phi, \phi^\dagger) = \int ds \left(i \phi^\dagger \partial_s \phi - \phi^\dagger K\phi\right).
\end{equation}
This formulation (\ref{pathcmps}) of the one-dimensional cMPS state as a path integral is a natural guiding expression for the generalisation to higher-dimensional scenarios which we describe later (The fermionic case is identical, except that $\phi$ is now a vector of Grassmann fields. Of course, both the bosonic and fermionic cases yield exactly the same physical state $|\chi\rangle$, since they coincide on the 1-particle sector). Notice that we've dropped the limits from the integrals; the expression (\ref{pathcmps}) makes equal sense for quantum systems on $[0,l]$ as for the infinite case $(-\infty, \infty)$.

While the use of the auxiliary Fock space and its 1-particle sector to encode the virtual process might seem initially excessive in the one-dimensional scenario, it turns out to be much more flexible in the higher dimensional generalizations. There the auxiliary field system has genuine spatial extent, and permits generalizations that are not simply 1-particle sector restrictions. The degree to which extending off the 1-particle sector in these models brings new physics and deviates from the discrete tensor network description is at present unclear, and demands further investigation.

\subsection{Interpretation of the cMPS path integral}

The expression (\ref{eq:cmps2}) admits a simple, yet useful interpretation. A cMPS is a superposition of coherent states $|\Phi\rangle$ with some weighting $e^{iS}$ determined by the virtual dynamics. The standard intuition concerning coherent states is that they are the ``most classical'' states of a quantum system due to their saturation of the Heisenberg uncertainty relation. Thus, (\ref{pathcmps}) tells us that a cMPS is a superposition of ``classical'' field states centred around classical field configurations $\Phi:\mathbb{R}\rightarrow \mathbb{C}$ in phase space. These field configurations $\Phi$ themselves are scalar functions of a vector of \emph{auxiliary} classical fields $\phi:\mathbb{R}\rightarrow \mathbb{C}$. By interpreting that spatial variable $s$ as a temporal variable one can understand the action $S$ for these auxiliary fields as that of a $(0+1)$-dimensional quantum field, i.e., ordinary quantum mechanics. 

One therefore has the picture of an auxiliary system undergoing a classical trajectory of its discrete variables, however to gain information (by measurement) about a dynamically evolving quantum system we inevitable disturb it because of the back-action of the quantum measurement. The closest representation of the dynamics in this quantum setting is to \emph{continuously monitor} the evolving auxiliary system with a sequence of infinitesimally weak measurements \cite{caves:1987a}. By exploiting von Neumann's prescription for quantum measurement this process is then understood as \emph{entangling} the auxiliary system and an infinite collection of \emph{meter} systems. The combined auxiliary system+meter collection undergoes completely positive dynamics. In the continuum limit the meter systems constitute a quantum field with one extra spatial dimension, the reduced state of the meters alone is a quantum state. The cMPS coherent field state is then an imprint of the discrete trajectory, and is as classical a record as possible. The stength and manner of this imprint is entirely contained in the particular coupling $R(t)$. Each trajectory for the auxiliary system contributes a coherent field state, and the cMPS is simply a superposition of ``classical'' trajectories with the according weighting by the action $S$. 

\begin{figure}
\begin{center}
\includegraphics{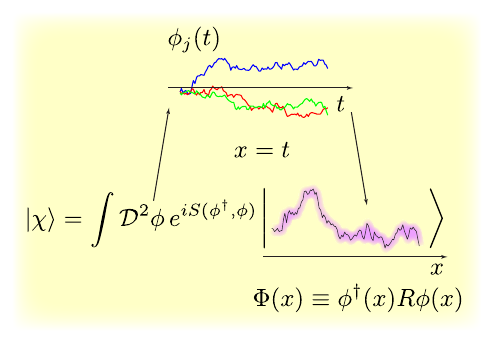}
\end{center}
\caption{\textbf{Interpretation of a continuous tensor network.} An illustration of the coherent-state path integral representation for a cMPS state $|\chi\>$: here a sample classical trajectory for the (in this case, three) auxiliary fields is depicted above. These classical trajectories are then combined via $\phi^\dag(x)R\phi(x)$ into a single complex scalar trajectory $\Phi(x)$. The field coherent state is then represented via the ket with the purple smeared trajectory, where a field coherent state is effectively a smeared-out classical configuration centred on $\varphi(x) \propto \rm{Im}(\Phi(x))$. The formula for the resulting cMPS is then a superposition of such coherent states weighted by the virtual action $S$.}
\end{figure}

\subsection{Completeness of the cMPS class}
In this section we show that the cMPS is a complete class: an arbitrary quantum field state can be approximated with arbitrary accuracy, by allowing the bond dimension $D$ to become arbitrarily large. The argument we present here is for the case of bosonic Fock space $\mathfrak{F}_-(L^2([0,l]))$ on a finite interval $[0,l]$ --  one expects this to hold in the case of the interval $(-\infty, \infty)$. 

The argument is rather simple and relies on three facts. The first is that an arbitrary quantum field coherent state 
\begin{equation}
	|\Phi\> \equiv \exp \left[\int_0^l \Phi(s) \sys^\dagger(s) - \Phi^*(s) \sys (s) \, ds \right] |\Omega_\A\>,
\end{equation}
is exactly representible as a cMPS $|\chi(\Phi)\rangle$ with bond dimension $D=1$. This follows upon taking $K$ and $R$ to be the one-dimensional matrices $K(s) = 0$ and $R(s) = \Phi(s)$. The boundary vectors $|\omega_L\rangle$ and $|\omega_R\rangle$ are simply taken to be equal to $1$. The next fact we require is that the span of all field coherent states is dense in Fock space, meaning that an arbitrary field state $|\Psi\rangle \in \mathfrak{F}_-(L^2([0,l]))$ may be approximated arbitrarily well by an increasing linear combination of field coherent states:
\begin{equation}
	\hspace{2cm}\sum_{l=0}^N c_j|\Phi_j\rangle \stackrel{N\rightarrow \infty} \longrightarrow|\Psi\rangle.
\end{equation}
This property follows from the over-completeness of coherent states in spanning the Hilbert space \cite{coherent}.The final fact we need is that a linear combination $|\chi\rangle = c_1|\chi_1\rangle + c_2|\chi_2\rangle$ of two cMPS $|\chi_1\rangle$ and $|\chi_2\rangle$ with bond dimensions $D_1$ and $D_2$, respectively, is again a cMPS with bond dimension $D=D_1+D_2$ and parameters $K = K_1\oplus K_2$, $R = R_1\oplus R_2$, $\langle \omega_L| = (c_1 \langle\omega_{L,1}|\oplus c_2\langle \omega_{L,2}|)$, and $|\omega_R\rangle = |\omega_{R,1}\rangle \oplus |\omega_{R,2}\rangle$.

Putting these facts together allows us to deduce that $\{|\chi_N\rangle \equiv \sum_{l=0}^N c_j|\Phi_j\rangle \} _N$ is a sequence of cMPS with bond dimensions $D_N = N$ that tend, in the limit, to an arbitrary state $|\Psi\rangle$ in Fock space. Thus we have confirmed the \emph{completeness} or \emph{expressiveness} property of the cMPS variational class in one dimension. It is worth noting that the argument we present here is by no means the most economical: there are, exploiting gauge invariance, almost certainly more efficient sequences of representations tending to the state $|\Psi\rangle$ using lower bond dimensions. Indeed, as we argue in the next section, a more economical representation of a physical field state is strongly suggested by the path integral representation.

It is worth noting that in the previous subsection we showed that an arbitrary cMPS is a superposition of field coherent states. Here we've shown the converse: an arbitrary superposition of field coherent states is also a cMPS.

\section{A toy example: cMPS representation of the final states of physical field dynamics.}

We can simply illustrate the previous components through the example of efficiently representing elementary field dynamics via cMPS.  Intuitively, we simply exchange the role of space and time and contract the (continuum) tensor network for the state of a dynamically evolving field along the spatial axis first, regarding it as a temporal axis. Beyond being a simple illustration of the present discussion, this construction also related to recent analysis of spin-systems \cite{2014arXiv1408.5140Z, 2014arXiv1411.2607R}. The generic situation we consider is therefore that of a bosonic field $\widehat{\psi}(x)$ in $\mathbb{R}$ obeying the canonical commutation relations
$	[\widehat{\psi}(x), \widehat{\psi}^\dag(y)] = \delta(x-y), $
with all other commutators vanishing. 

We can simply consider a second quantised hamiltonian with kinetic+potential energy split $\widehat{H} = \widehat{T} + \widehat{V} + \widehat{W}$, where
\begin{equation}
	\widehat{T} = \int \frac{d\widehat{\psi}^\dag(x)}{dx}\frac{d\widehat{\psi}(x)}{dx}\, dx, \hspace{0.3cm}
	\widehat{V} = \int V(x)\widehat{\psi}^\dag(x)\widehat{\psi}(x) \, dx,
\end{equation}
and with an interaction potential
\begin{equation}
	\widehat{W} = \int W(x-y) \widehat{\psi}^\dag(y)\widehat{\psi}^\dag(x)\widehat{\psi}(x) \widehat{\psi}(y) \, dxdy.
\end{equation}
For the sake of illustration, it suffices to concentrate on pointlike interactions, i.e. $W(x-y) = w\delta(x-y)$, with $w$ a constant. We then consider a field system, initialized in the physical state $|\varphi (0) \> \in \H_\A$ and allow it to evolve under the full hamiltonian $\widehat{H}$ for a time $T$, until it reaches the state $|\varphi (T)\> = e^{-iT\widehat{H}} |\varphi (0)\>$. Our task here is to illustrate how the final state $|\varphi (T)\>$ can be described in terms of the cMPS path integral representation, which can be interpreted instead as a (virtual) process in which some additional auxiliary system $\H_\B$ undergoes dissipative dynamics that couple it to the physical field system and on completion generates $|\varphi(T)\>$. To avoid confusion, in the case of the \emph{physical} evolution of the field system $\H_\A$ we use $x$ for the spatial coordinate, and $t$ for the physical time coordinate, while for the virtual process in which the auxiliary system $\H_\B$ couples to the physical field $\H_\A$ we use $s$ for the virtual time coordinate of $\H_\B$ and label \emph{subsystems} of $\H_\B$ with the parameter $\beta$. The construction that follows will roughly amount to reinterpreting the field variables $(x,t)$ as $(s,\beta)$ within a quite physically distinct setting.

The first move is to reformulate the physical field evolution in terms of a path integral expression over coherent states. The construction proceeds, through a Trotter-discretization of the time interval $[0, T]$ into $n$ pieces of length $\epsilon = T/n$ and writing
\begin{equation}
	|\varphi(T)\rangle = (e^{-i\epsilon\widehat{H}})^n |\varphi(0)\rangle.
\end{equation} 
We suppose, for simplicity, that the initial state $|\varphi(0)\rangle$ is a coherent state.

As in the construction of the auxiliary action, we insert a resolution of the identity, in terms of 1-d field coherent states $|\Phi(t)\> :=\exp [\int \! dx  \,\Phi(x,t) \widehat{\psi}^\dagger(x) - \Phi^*(x,t) \widehat{\psi} (x)] |\Omega\>$, between each application of $e^{-i\epsilon\widehat{H}}$. 
Expanding up to first order, and using the overlap equation (\ref{coherentoverlap}) we have that an infinitesimal advance for the physical system is described by

\begin{equation}
	\hspace{-1cm}\langle \Phi(t)|  e^{-i\epsilon\widehat{H}} |\Phi(t-\epsilon) \rangle \approx e^{-\frac{\epsilon}{2}\int {\Phi^*(x,t)}\partial_t\Phi(x,t)  - \partial_t{\Phi^*(x,t)}\Phi(x,t)     -i\epsilon \H(\Phi^*(x,t), \Phi(x,t))\, dx},
\end{equation}
with a hamiltonian density $\H(x,t)$ given by
\begin{equation}
	\H(\Phi^*(x,t), \Phi(x,t )) = |\partial_x \Phi (x,t)|^2 + \ V(x)|\Phi (x,t)|^2+  w |\Phi (x,t)|^4  
\end{equation}
Summing over each time interval yields
\begin{equation}\label{eq:realtime}
	|\varphi(T)\rangle = \int \mathcal{D}\Phi \mathcal{D}\Phi^* e^{iS(\Phi, \Phi^*)}|\Phi(T)\rangle,
\end{equation}
being a superposition of physical coherent states described by $\Phi(x,T)$ at time $t=T$, and with the action
\begin{equation}
	S(\Phi, \Phi^*) = \int_0^T \int_{-\infty}^\infty  i{\Phi^*(x,t)}\partial_t\Phi(x,t) - \H(\Phi^*(x,t), \Phi(x,t)) \, dxdt.
\end{equation}
The lower limit of this path integral is $\Phi (x,0) = \varphi(x,0)$ while the upper limit is unconstrained.

The path integral form of $|\varphi (T)\>$ is suggestive of how an auxiliary system should couple to the physical system in order to generate $|\varphi (T)\>$ under (virtual) dissipative dynamics. Since we wish the auxilary system $\H_\B$ to sweep over the length of the physical field the time parameter for the process $s$, should correspond to the physical spatial variable $x$.

To capture this idea we can subdivide the auxiliary system $\H_\B$ into harmonic oscillator subsystems as $\H_\B = \otimes_\beta \H_\beta$, labelled by some variable $\beta$, however since $t$ is a continuous variable we effectively take the limit in which $\H_\B$ is an \emph{auxiliary} complex field where $\beta$ is its spatial coordinate and the auxiliary system has spatial extent from $\beta = 0$ to $ \beta =T$. The key point is that spatial couplings (along $\beta$) within the hamiltonian of the auxiliary system can be used to simulate the physical dynamics that generates $|\varphi (T)\>$, as the auxiliary system sweeps out over the physical field, and couples to it through a natural interaction term.

For the auxiliary variables we use $\widehat{z}_0(s,\beta)$ and $\widehat{z}_1(s,\beta)$, which we can combine into a single complex field as $\widehat{z}= \widehat{z}_0 + i\widehat{z}_1$. The hamiltonian of the auxiliary system is taken to be
\begin{eqnarray}\label{freeham}
	\hspace{-2cm}\widehat{K} (s) = \int_0^Td \beta  [  -\frac{1}{4}\widehat{p}_0(s,\beta)^2 - \frac{1}{4}\widehat{p}_1(s,\beta)^2 + V(\widehat{z}_0(s,\beta)^2+\widehat{z}_1(s,\beta)^2)  + w(\widehat{z}_0(s,\beta)^2\nonumber\\
	+\widehat{z}_1(s,\beta)^2)^2  
- i(\widehat{z}_0(s,\beta)-i\widehat{z}_1(s,\beta))\partial_\beta (\widehat{z}_0(s,\beta)+i\widehat{z}_1(s,\beta)) ],
\end{eqnarray}
with $\widehat{p}_0$ and $\widehat{p}_1$ the momenta conjugate to $\widehat{z}_0$ and $\widehat{z}_1$.

The form of (\ref{eq:realtime}) suggests that the interaction term coupling the auxiliary and physical systems be taken to be the continuous measurement interaction in which the physical system $\H_\A$ is interpreted as continuously measuring the `observable' $\widehat{z} = \widehat{z}_0 + i\widehat{z}_1$. This is obtained as the continuum limit of the coupling 
\begin{equation}\label{interaction}
	\widehat{H}_{\mathrm{int}}(s) =  i\epsilon\sum_{j\in \mathbb{Z}} \delta(s-j\epsilon) \left[ \widehat{z}(s,\beta=T)\otimes \widehat{\psi}^\dag_{j_\A} - \widehat{z}^\dagger(s,\beta=T)\otimes \widehat{\psi}_{j_\A} \right],
\end{equation}
in other words, the physical system only couples to the extreme edge of the auxiliary system at the (auxiliary) spatial point $\beta=T$. Here $\widehat{\psi}_{j_\A} \equiv \frac{a_{j_\A}}{\sqrt{\epsilon}},$ and $a_{j_\A}$ is the operator which annihilates a boson with wavefunction $\frac{1}{\sqrt{\epsilon}}\chi_{[(j-1)\epsilon, j\epsilon)}(x)$ for the physical system.

It is now a case of checking that the composite system $\H_\A\otimes \H_\B$, evolving under the full hamiltonian $\widehat{H}_{\mathrm{tot}} = \widehat{K}+\widehat{H}_{\mathrm{int}}$ for auxiliary time from $s=-\infty $ to $s=+\infty$ will indeed generate the desired field state $|\varphi (T)\>$ as expressed in the path integral form (\ref{eq:realtime}). The calculation proceeds in a similar manner to the earlier cMPS path integral calculation evolving under the composite hamiltonian $\widehat{H}_{\mathrm{tot}}$, however for our resolution of the identity at auxiliary time $s$ we use the complete set of states $\{|z(s)\>\}$ given by
\begin{eqnarray}
|z(s)\> = |z_0(s,0), z_0(s,\epsilon), \cdots z_0 (s, T) ; z_1(s,\epsilon), z_1(s,2\epsilon), \cdots z_1 (s, T) \>,
\end{eqnarray}
which we express in the discretized setting with oscillators located at $\beta=0,\epsilon, 2\epsilon,\dots, T$. A straightfoward calculation gives that
\begin{equation}
\hspace{-2cm}\<\omega_L|\mathcal{P}e^{-i\int{\widehat{H}_{\mathrm{tot}}(s)ds}}|\omega_R\> |\Omega\>
= \int \mathcal{D}^2z(s,\beta) \mathcal{D}^2p(s,\beta)\ \exp{[iS'(p_0,p_1,z_0,z_1)]} |\Omega_\A\>
\end{equation}
where we have the action
\begin{equation}
\hspace{-2cm}S' = \int_{-\infty}^\infty ds\int_{0}^Td\beta  \, (p_0\dot{z}_0 + p_1\dot{z}_1 - K(p,z) - iz(s,\beta=T)\widehat{\psi}^{\dagger}_\A + iz^*(s,\beta=T)\widehat{\psi}_\A).
\end{equation}
Consequently, by integrating out $p_0$ and $p_1$, and identifying $z(s, \beta=T)$ with $\Phi(x,T)$ we see that the evolved physical state $|\varphi(T)\>$ can be represented by a cMPS with free hamiltonian $K$ given by (\ref{freeham}) and interaction given by (\ref{interaction}). We should emphasize that it is the final state for which we are providing an efficient description, and not the dynamics. The dynamics that generated the state is easily calculated, and acts to ensure that such a description exists, however the representation via virtual dissipative dynamics can go beyond such cases and can provide novel tools for non-trivial quantum states, such as the ground state of strongly interacting systems.

The cMPS representation that we have constructed involves an infinite dimensional auxiliary system where integration over $\beta$ corresponds to a continuum summation over the auxiliary indices; this is not unexpected since the auxiliary system faithfully simulates the entire dynamical history of the physical field. However, the local character of the interaction term implies that we can obtain $|\varphi (T)\>$ equally well from the coupling of a single auxiliary oscillator to the physical field, with the composite system now undergoing a more general completely-positive map (instead of a unitary interaction). Specifically, the above calculation has shown that $|\varphi (T)\> = \< \omega_L | U |\omega_R\> |\Omega\>$, or more generally $|\varphi (T)\>\< \varphi (T)| = \Tr_{\mathrm{aux}} [ U (\omega \otimes |\Omega\>\<\Omega| )U^\dagger] $ for some operator $U$ on the joint system and auxiliary state $\omega$, but which can now be written as $\Tr _{\beta=T}[\, \Tr_{\beta \ne T} [U (\omega \otimes |\Omega\>\<\Omega| )U^\dagger]]  = \Tr_{\beta=T} [ \mathcal{E} (\omega_\beta \otimes |\Omega\>\<\Omega| )]$ for some completely-positive map $\mathcal{E}$ defined on the physical field and oscillator at $\beta=T$. By truncation of the oscillator Hilbert space, and simulation of the evolution $\mathcal{E}$ we may thus obtain an efficient cMPS description of $|\varphi(T)\>$ in terms of a purely discrete auxiliary quantum system.

\section{Beyond one-dimension: The continuum limit of a PEPS class}\label{sec:cpeps}

We have seen the one can represent a general cMPS via the path integral over an auxiliary $(0+1)$-dimensional  $D$ component complex field $\phi$ where the path integral is a coherent-state path integral over all configurations of the $D$-dimensional complex vector $\phi$, and where the auxiliary system is subject to an action $S(\phi, \phi^\dagger) = \int ds \left(i \phi^\dagger \partial_s \phi - \phi^\dagger K\phi\right)$.

Such a representation of $|\chi\rangle$ naturally suggests a higher dimensional generalisation, namely, we should simply have that 
\begin{equation}\label{2dpathcmps}
|\chi\> = \int \D^2\phi\, \exp \left[ -S (\phi, \phi^\dagger) \right] |\Phi\>,
\end{equation}
where the path integral is now over an auxiliary $(d-1+1)$-dimensional  field $\phi(\mathbf{z},t)$ with $D$ components, $|\Phi\rangle$ is now a higher-dimensional field coherent state 
\begin{equation}
	|\Phi\> \equiv \exp \left[\int \Phi(\mathbf{x}) \sys^\dagger(\mathbf{x}) - \Phi^*(\mathbf{x}) \sys (\mathbf{x}) \, d\mathbf{x} \right] |\Omega_\A\>,
\end{equation}
where $\Phi(\mathbf{x}) = \phi^\dagger(\mathbf{x}) R\phi(\mathbf{x})$, $R$ is a $D\times D$ matrix, and  $S$ is a local complex action for a $D$ component auxiliary boundary field $\phi$ living on an auxiliary boundary space of one lower dimension. Note the notation $\phi(\mathbf{x})$ denotes $\phi$ at $(\mathbf{z},t)$ via regarding the first $d-1$ components of $\mathbf{x}$ as spatial coordinates and the $d$th component as a temporal coordinate, i.e., $z_j = x_j$, $j=0, 1, \ldots, d-2$, and $t = x_{d-1}$. Also note that we adopt a euclideanised action, a point which will later prove advantageous when imposing symmetries on the physical state.

%%%%%%
While taking the continuum limit of the one-dimensional MPS class is comparatively straightforward, the two-dimensional equivalent poses more problems. It is true that we can simply posit the form of a two-dimensional (or higher-dimensional) cMPS as being generated by the continuous measurement process of a lower-dimensional auxiliary boundary field \cite{osborne:2010a}, however this is unsatisfactory for at least two reasons. Firstly, in such a setting it is not clear, a priori, how one might impose certain desirable symmetries, such as rotational symmetry, on the physical quantum state. Any variational class intended for the efficient description of real-world physics should be capable of manifestly exhibiting such symmetries. Secondly, for discrete systems higher-dimensional generalizations of MPS already exist, such as the PEPS class, which have been powerful tools in understanding the physics of local hamiltonians. As such it is also of theoretical importance that we arrive at a continuum limit of PEPS that mirrors the one-dimensional cMPS class. 

In previous sections we obtained a path integral representation for the one-dimensional cMPS class from the traditional discrete MPS class by taking a well-behaved continuum limit, and which we can use as our guide for constructing higher-dimensional classes with manifest symmetries; in field theory, path integral formulations are ideally suited for the imposition of symmetries that would not be manifest according to, e.g., canonical quantisation of the field. Our strategy is then to develop a continuum limit as a superposition of field coherent states with amplitudes given by a path integral over an auxiliary system and such that desirable symmetries are manifest.

\subsection{The basic Tensor Network setting beyond 1-d systems.}
A natural higher-dimensional generalization of MPS are the Projected Entangled Pair states (PEPS), which are examples of \emph{Tensor Networks} \cite{verstraete:2008review, murg:2007,Verstraete:cond-mat0407066}. The original formulation of PEPS rested on distributing maximally entangled pairs of $D$-dimensional quantum systems between neighbouring sites on a graph, and then locally mapping the systems at each point into a single $d$-dimensional Hilbert space. The PEPS construction for arbitrary $D$ can describe any quantum state, and is naturally suited to systems displaying area laws. A generic PEPS has an expansion in terms of a product basis with expansion coefficients given by a contraction of tensors $A^r_{(i \cdots k)}$ with respect to a particular graph $\Gamma (V,E)$:
 \begin{eqnarray}\label{peps}
|\chi\> &=& \C [ A^{r_1}_{(i_1\cdots k_1)} \cdots A \cdots A^{r_N}_{(i_N \cdots k_N)}] | r_1 \cdots r_N \>,
\end{eqnarray}
where $\C$ denotes a complete contraction of the auxiliary indices  $(i \cdots k)$ according to the graph edge structure $E$, and $r_1, \dots r_N$ label the (product) configurations of a the discrete physical systems located at each vertex of the graph.

An initial instinct would be to begin with a two-dimensional square lattice, and embed the discrete system into the one-particle sector of a system of bosonic or fermionic auxiliary fields, as was done previously for the 1-D cMPS path integral. If one directly follows this path, passing from the discrete PEPS to a continuum path integral, one finds that the underlying square lattice structure persists in the field, and one does not obtain a rotation invariant physical state (see \ref{squarelatticeappendix}). Here we adopt a slightly more involved strategy to handle this unwanted feature.

We begin with a graph of $N^2$ points $\{ (x,y)= (n\epsilon,m\epsilon) : n,m \hspace{0.2cm} \rm{ integers} \}$, with a physical spacing $\epsilon$. We view the $y$ spatial direction as an auxiliary time $t$.
%, and auxiliary states $|\omega\>$ defined over the $x$ direction. 
As written (\ref{peps}) involves a contraction over $N^2$ auxiliary subsystems distributed over the graph, with independent couplings to the physical degrees of freedom at each site. Since we wish to view the $y$ direction as an auxiliary time dimension we regard the contraction over the $N^2$ subsystems as the sum over configurations of $N$ auxiliary subsystems subject to a sequence of $N$ dynamic transformations. The upshot is to replace the contraction along the $y$ direction with a product of $N$ square matrices of dimension $D^N$. This is well-known as a \emph{transfer operator} approach, i.e., we simply view the contraction from one value of $y$ to the next as multiplication by a particular transfer operator. The state (\ref{peps}) can then be written as 
\begin{equation}
|\chi\rangle = {}_\mathcal{B}\<\omega_L |\widehat{U}(t=l,t=0)|\omega_R\>_{\mathcal{B}}|\Omega\rangle_\mathcal{A},
\end{equation} 
where $\widehat{U}$ is an operator on $\mathcal{AB}$ given by the (time-ordered) product of transfer operators local with respect to the graph sites and $|\Omega\rangle_{\mathcal{A}}$ is some initial product state of $\mathcal{A}$. Specifically, $\widehat{U} =  \T \prod_{t=0 }^l[ \widehat{M}(t)] \equiv \widehat{M}(l)\widehat{M}(l-\epsilon)\widehat{M}(l-2\epsilon)\cdots \widehat{M}(0)$, where the transfer operator $\widehat{M}(t)$ generates an elementary time-step of size $\epsilon$ and is built from local operators on $\mathcal{A}$ and $\B$. We then follow the idea used for the 1-dimensional case and regard each the auxiliary system at lattice site of $\mathcal{B}$ as the single-particle space of the Fock space built from $\mathbb{C}^D$. 

The contraction of indices depends on the particular graph structure being used. However, our goal is to construct cMPS states with symmetries and we follow the key principle that \textit{the symmetries of the physical state are encoded in the dynamics of the auxiliary system}. For example, a natural symmetry to demand is that of rotation invariance in the spatial coordinates of the physical field state. Assuming a state of the form (\ref{2dpathcmps}) implies that the auxiliary action $S$ is invariant under the induced $\textsl{SO}(2)$ rotation group (assuming the measure is also invariant). By demanding that the auxiliary system is a physical system we deduce that $S$ should be an action describing the completely positive dynamics auxiliary system (after we trace out $\mathcal{A}$). However, encoding the symmetry into the dynamics of the auxiliary system means imposing invariance under $\textsl{SO}(2)$. This implies that the dynamics should be viewed as the imaginary-time evolution of a Lorentz-invariant system (which is still a completely positive map of the quantum state). It is also useful to emphasize that technical subtleties arise when taking the limit of lattice systems. Specifically, we might consider a family of graphs $\{\Lambda_k \}$ indexed by some variable $k=0,1,2,\dots$, that converges to some dense subset of a compact spatial region $A$. To each point $\mathbf{x} \in \Lambda_k$ we have an associated Hilbert space $\H (\mathbf{x})$, which could be a space of finite or countably infinite dimension. The total Hilbert space for the full graph system is then given by $\H_k=\otimes_{\mathbf{x}\in \Lambda_k} \H(\mathbf{x})$, and in the thermodynamic limit $k\rightarrow \infty$, the resultant space will have an uncountable dimension. One instead works with a much smaller, separable Fock space $\mathfrak{F}(\H)$ constructed to ensure that every state in $\mathfrak{F}(\H)$ has finite particle expectation value, and splits up into a sum $\mathfrak{F}(\H)=\oplus_n \H^{(n)} $ of particle sectors $\H^{(n)}$ with finite particle numbers. Central to the formation of this Hilbert space is the identification of a vacuum state, from which the different $n$-particle spaces $\H^{(n)}$ are obtained through the action of creation operators obeying the desired statistics. It is well-known that the Stone-von Neumann theorem fails for these systems, and many unitarily inequivalent Fock spaces may be constructed through the choice of different vacua and creation/annihilation operators. For our analysis of the discrete to continuum limit, we specify the local Hilbert spaces at each point on the graph, but ultimately we make use of a Fock space construction for the state $|\chi\>\in \mathfrak{F}(\H_\A)$, and work with a particular choice of creation/annihilation operators for both $\A$ and $\B$, with the auxiliary system $\B$ carrying bosonic or fermionic statistics.
 
Our strategy is then to first construct a Lorentz-invariant auxiliary action from the continuum limit of a sequence of discrete PEPS. We then construct an analytic continuation to the Euclidean setting and obtain a one-parameter family of discrete PEPS states giving a rotation invariant Euclidean action as $\epsilon \rightarrow 0$. There are clearly different possible choices for a Lorentz-invariant action; motivated by the first-order action (\ref{cmpsaction}), and certain convenient properties of coherent field states, we derive a Dirac-like action from a specific sequence of PEPS. One might question why we bother going via a Lorentz-invariant setting. The reason is that if we begin with $SO(2)$ symmetry as our target then we do not have ready access to the intuition that the physical field state is generated by the virtual dynamics of a lower dimensional system.  

The first task is to arrive at a Lorentz-invariant situation, a problem for which physical intuition is readily available. Since we are looking for an  auxiliary $(1+1)$-dimensional lattice system with locally defined dynamics we assume that each site $(x,t)$ has contraction links to future sites $(x,t+\epsilon), (x-\epsilon, t+\epsilon)$ and $(x+\epsilon, t+\epsilon)$ and also to past sites $(x,t-\epsilon), (x-\epsilon, t-\epsilon)$ and $(x+\epsilon, t-\epsilon)$. The simplest such choice is to build the operator $\widehat{M}$ out of quadratic terms involving creation and annihilation operators; to arrive at states with rotational symmetries, we can also make use of spinorial expressions. To generate the spinorial structure we assume that at each site $x$, in addition to the `flavour' indices $i,j,k,\dots$, we have access to two internal degrees of freedom, with annihilation operators $a_{k,x}$ and $b_{k,x}$ at each spacetime point. We also note that the bosonic and fermionic cases can be treated simultaneously by being careful with the ordering of terms. Thus we have $[ a_{j,x} , a_{k,y}]_\pm =[ a^\dagger_{j,x} , a^\dagger_{k,y}]_\pm = 0, $ $[ a_{j,x} , a^\dagger_{k,y}]_\pm = \delta_{jk} \delta_{x,y}$, with similar expressions for $b$, where $\pm$ labels the choice of bosonic or fermionic auxiliary system.

\begin{figure}[h]
\begin{equation*}
	\hspace{-0.8cm}\xymatrix{
	 &(x-\epsilon, t+\epsilon) &(x,t+\epsilon) &(x+\epsilon, t+\epsilon) &(x+2\epsilon,t+\epsilon) & \\
	  &*+[F:<3pt>]\txt{$\H_{ab}(x-\epsilon,t)$}\ar[d]\ar[dl]\ar[dr]\ar[ul]\ar[ur]\ar[u]&*+[F:<3pt>]\txt{$\H_{ab} (x,t)$}\ar[d]\ar[dl]\ar[dr]\ar[ul]\ar[ur]\ar[u]&*+[F:<3pt>]\txt{$\H_{ab}(x+\epsilon,t)$}\ar[d]\ar[dl]\ar[dr]\ar[ul]\ar[ur]\ar[u]&*+[F:<3pt>]\txt{$\H_{ab}(x+2\epsilon,t)$}\ar[d]\ar[dl]\ar[dr]\ar[ul]\ar[ur]\ar[u]  &\\
&(x-\epsilon, t-\epsilon)&(x,t-\epsilon)&(x+\epsilon, t-\epsilon)&(x+2\epsilon,t-\epsilon)&
}
\end{equation*}
\caption{Tensor network structure for the transfer operator generating one temporal layer of the PEPS sequence: here $\H_{ab}$ denotes the tensor product $\H_a\otimes \H_b$ for the internal degrees of freedom of the auxiliary system $\H_\B$.}\label{mhatOp}
\end{figure}
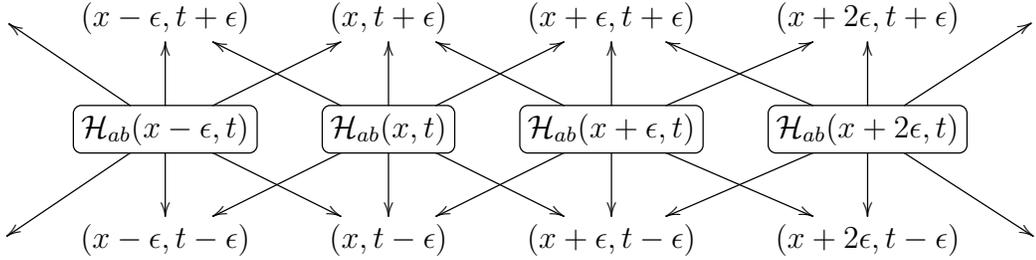 

\subsection{Generation of a kinetic term and flavor doubling}
We now construct a transfer operator $\widehat{M}_\epsilon = \I + \epsilon \widehat{H}_{\rm tot}$ from terms quadratic in $a$ and $b$, and generate a PEPS via
\begin{equation}
	|\chi_\epsilon\rangle = {}_\mathcal{B}\<\omega_L |\widehat{M}_\epsilon(l)\widehat{M}_\epsilon(l-\epsilon)\widehat{M}_\epsilon(l-2\epsilon)\cdots \widehat{M}_\epsilon(0)|\omega_R\>_{\mathcal{B}}|\Omega\rangle_\mathcal{A},
\end{equation}
in such a manner that the continuum limit has the desired Lorentz symmetry. It should perhaps be emphasized that the generators that we construct relate entirely to the \emph{auxiliary} system $\B$, which only forms an abstract device to describe the physical state $|\chi_\epsilon \>$, and does not commit us to any particular realisation for the two-dimensional \emph{physical} field system $\A$. However, by treating the auxiliary system as physical we can make use of natural intuitions of particle interactions when we construct the abstract PEPS class through $\widehat{H}_{\rm tot}$. Certain assumptions are natural to impose on the terms appearing in $\widehat{H}_{\rm tot}$, such as left-right symmetry, symmetry between a-particles and b-particles and conservation of total particle number, however the key term in the construction is the a nearest neighbour `hopping' term, which we take to be
\begin{equation}
H_{\rm h}(t) = \frac{1}{\epsilon}\sum_x J^{jk}(t)(a^\dagger_{j,x} b_{k,x-\epsilon} +a^\dagger_{j,x} b_{k,x} + a^\dagger_{j,x} b_{k,x+\epsilon}+ \rm{h.c.}),
\end{equation}
where $x$ runs over $N$ spatial points, and $J^{jk}(t)$ measures the strength of the spatial hopping, which for simplicity we take as constant along the spatial direction.

To analyse this, we perform a discrete Fourier transform in the spatial direction to obtain
\begin{equation}
H_{h}= \frac{1 }{\epsilon}\sum_{p,j,k}J^{jk}(t)
(1+2\cos p\epsilon )(\tilde{a}_{j,p}^\dagger \tilde{b}_{k,p} +\tilde{b}_{j,p}^\dagger \tilde{a}_{k,p} ),
\end{equation}
where $\tilde{a}_{j,p} = \frac{1}{\sqrt{N}} \sum_x e^{-ipx} a_{j,x}$ is the Fourier-transformed annihilation operator (similarly for $\tilde{b}_{j,p}$), and $p=2\pi n/N\epsilon$ runs over $N$ points in the reciprocal lattice, for $n$ an integer.

Our concern is the continuum limit, $\epsilon \rightarrow 0$, where the dominant contributions of $H_h$ will come from the ``low-energy regime'' of momenta $p$ near to the zeroes of $(1+2 \cos p \epsilon)$. These occur at the points $q_\mu =(-1)^\mu(2\pi/3\epsilon)$, for $\mu=0,1$.  The contributions from the two points give rise to two flavors in a similar manner to fermionic doubling on the lattice, however it must be noted that this doubling only occurs for the auxiliary system, and so the physical system (whether bosonic or fermionic ) is unaffected. Overall, in the continuum limit $H_h$ splits into $H_h = H_{h,0} + H_{h,1}$ with the contributions from the two decoupled flavors given by
\begin{equation}
H_{h,\mu}(t) = \sqrt{3}\sum_{ |p|\ll 1/\epsilon  }(-1)^\mu J^{jk}(t) \left ( \begin{array}{lr} \tilde{a}_{j,q_\mu + p}^\dagger  \tilde{b}_{j,q_\mu+ p}^\dagger \end{array}\right ) p \sigma_x \left (\begin{array}{c} \tilde{a}_{j,q_\mu +p}\\  \tilde{b}_{j,q_\mu+ p}\end{array}\right )
\end{equation}
We can redefine $J^{jk} \rightarrow J^{jk}/\sqrt{3}$ and relabel the mode operators as $\tilde{a}_{(j, \mu) ; p} \equiv \tilde{a}_{j, q_\mu+p}$ and also $\tilde{b}_{(j, \mu) ; p} \equiv \tilde{b}_{j, q_\mu+p}$ to obtain
\begin{equation}
H_{h}(t) =  \!\!\!\!\! \sum_{ |p| \ll 1/\epsilon ; \mu=0,1 }\!\!\!J^{jk}(t) \left ( \begin{array}{c} \tilde{a}_{(j,\mu), p}^\dagger  \tilde{b}_{(j,\mu) ,p}^\dagger \end{array}\right ) p \sigma_z^\mu\sigma_x \sigma_z^\mu \left (\begin{array}{c} \tilde{a}_{(j, \mu), p}\\  \tilde{b}_{(j,\mu) ,p}\end{array}\right ).
\end{equation}
This term alone would produce $2D$ massless Lorentz-invariant flavors labeled by the compound index $(j,\mu)$, in which the $(j,0)$-field is related to the $(j,1)$-field through the discrete parity transformation $\P$, given in 1+1 dimensions as $\P = \gamma^0=\sigma_z$, which inverts chirality. This is consistent with the Nielsen-Ninomiya theorem \cite{Nielsen-Ninomiya:1981}, which requires doubling in order to achieve a translationally invariant spinor action with chiral symmetry in the continuum limit of a lattice model.

\subsection{Coupling and decoupling of the two flavors}
The previous analysis shows that the two contributions to $H_h$ from the points $q_0$ and $q_1$ in momentum space decouple. In position space this tells us that $a_{j,x}$ splits up in the low-energy regime as $a_{j,x} = \{ \sum_{|p| \ll 1/\epsilon} e^{ipx} \tilde{a}_{q_0+p}\} e^{iq_0x} + \{ \sum_{|p| \ll 1/\epsilon} e^{ipx} \tilde{a}_{q_1+p}\} e^{iq_1x}$. We write this instead as $a_{j,x} = a_{(j,0),x}e^{iq_0x} + a_{(j,1),x}e^{iq_1x}$, in which the operators $a_{(j,\mu),x}$ are given by the expressions in the curly brackets of the preceding equation.

In the continuum limit we then have that $\epsilon^{-1/2}a_{(j,\mu),x}$ tends to a \emph{smooth field} $\Psi_{(j,\mu)} (x)$, arising from envelopes of plane waves around the point $q_\mu $. This real-space description has been useful recently to generate non-trivial field potentials from lattice models \cite{Alba:2011}. For example, one might perturb $H_h$ through the addition of an on-site potential such as $\epsilon \sum_x f^{jk}(x) a^\dagger_{j,x} a_{k,x}$. This term will behave as $\epsilon \sum_{\mu , \nu,x} f^{jk}(x) e^{-i (q_\mu - q_\nu)x}a^\dagger_{(j,\mu),x}a_{(k,\nu),x}$, however, if the function $f^{jk}(x)$ does not vary rapidly from site to site, the presence of the highly oscillatory term  $e^{-i(q_\mu - q_\nu)x}$ will ensure that only $\mu=\nu$ will contribute in the continuum limit, and so the two flavors will decouple. At the other extreme, one can consider functions on the lattice that vary sufficiently rapidly, as say $f (x)= g(x(q_0-q_1))$, which can be used to produce non-trivial couplings of the flavors \cite{Alba:2011, Jackiw:2007}, however here we do not consider such scenarios.

\subsection{Generation of the full transfer operator}
It is straightforward to produce a mass term in the continuum limit, simply by adding the term $\sum_x m (a^\dagger_{j,x}a_{j,x}- b^\dagger_{j,x} b_{j,x})$, which then generates the usual dispersion relation $E^2 = p^2 +m^2$ as $\epsilon \rightarrow 0$. However, we can more generally use functions $\{ m_0^{jk}(x,t)\}$ that do not vary too rapidly over the lattice, and perturb $H_h$ by the on-site potential term $H_m (t)= \sum_x m_0^{jk} (x,t) ( a^\dagger_{j,x} a_{k,x} - b^\dagger_{j,x} b_{k,x})$. This can be written as 
\begin{equation*}
\hspace{-2.5cm}H_m(t) \!=\! \sum_{x,j,k,\mu }\left ( m^{jk}_0(x,t)( a^\dagger_{(j,\mu),x} a_{(k,\mu),x} - b^\dagger_{(j,\mu),x} b_{(k,\mu),x}) + e^{\pm i(q_0-q_1)x}\mbox{(flavor coupling terms}) \right )
\end{equation*}
where the flavor coupling terms do not contribute in the continuum limit, as explained in the previous section. In addition to the terms $H_h$ and $H_m$ for the auxiliary system we add a coupling term between the auxiliary and physical systems, which will generate the state $|\chi_\epsilon\>$. For this, we mirror the local coupling used for the 1-dimensional system and define the interaction term
\begin{equation}\label{Hint}
\widehat{H}_{\rm int} (t) =  \sum_x  R^{jk}(x,t) 
\left [ ( a^\dagger_{j,x} a_{k,x} + b^\dagger_{j,x} b_{k,x}) \otimes \sys^\dagger (x, y=t) \right ],
\end{equation}
where again, for simplicity, we assume the functions $\{ R^{jk}(x,t)\}$ vary sufficiently slowly over the lattice so that the $\mu=0,1$ flavors decouple once more.

The final transfer operator that generates the PEPS state is finally given by
\begin{equation}\label{eq:mtxop}
\widehat{M}_\epsilon= \I + \epsilon (H_{\rm m} + H_{\rm h} + \widehat{H}_{\rm int}),
\end{equation}
where the operator hat serves to specify the non-trivial action on the physical field system. The basic tensor structure of $\widehat{M}_\epsilon(t)$ is shown in Fig.~\ref{mhatOp}, where there is also an implicit physical index at each site, coupling to the physical field $\A$, which we omit in the diagram for the sake of clarity. 

In the next section we analyse the continuum limit, and derive the desired path integral representation from the smooth fields $\epsilon^{-1/2}a_{(j,\mu),x}$ and $\epsilon^{-1/2}b_{(j,\mu),x}$. On the assumption that $m_0^{jk}$ and $R^{jk}$ slowly vary on the lattice, the expressions for $H_m$ and $\widehat{H}_{\rm int}$ in terms of these smooth fields are obtained by the doubling of flavor index $j \rightarrow (j,\mu)$. The same happens for the indices of the kinetic term $H_h$, but with added feature of a parity flip $\sigma_x \leftrightarrow \sigma_z \sigma_x \sigma_z$ relating the two flavors. As we shall see later, since this discrete symmetry is itself a Lorentz symmetry the total field state that arises will possess the full symmetry group that we require.

\subsection{Construction of the path integral representation}
In this, and the subsequent, subsection we construct the path integral for the continuum limit of the sequence $|\chi_\epsilon\rangle$, $\epsilon\rightarrow 0$. We follow the one-dimensional prescription, and insert coherent-state resolutions of the identity into the product $\T \prod_{t=0}^l [\widehat{M}_\epsilon(t)]$. This is first computed for a simple elementary timestep $\epsilon$ and the continuum limit taken in the spatial direction. Finally the continuum limit is taken in the timelike direction, to obtain the final path integral representation of the cMPS state.

For clarity we shall write $j_\mu$ to denote the compound index $(j,\mu)$ with $j_0=1,\dots ,D$ and $j_1=1, \dots ,D$ for the two flavor sectors. This notation is useful since the actual PEPS parameters $m^{jk}_0$ and $R^{jk}$, that we can control do not have any $\mu$ dependence, and so the $\mu$ label simply doubles up the auxiliary fields, without playing an independent variational role. In what follows will use coherent-state resolutions of the identity which, in both the fermionic and bosonic cases, are given by
\begin{equation}
|\bm{\phi}(x,t)\> \equiv |\otimes_{j_\mu,x,s} \phi_{j_\mu ,s} (x,t)\> :=D(\{\phi_{j_\mu,s} (x,t)\}) |\Omega_\B\>,
\end{equation}
 where we have the usual coherent state displacement operator, given for a single mode with annihilation operator $c$ as $D(\alpha) = \exp[ \alpha c^\dagger - c \alpha^*]$, the label $s=a,b$ labels the particle type, and in the fermionic case $\phi_{j_\mu,s} (x,t)$ are Grassmann numbers. We also use bold-faced $\bm{\phi}:=\{\phi_{j_\mu,s} (x,t)\}$ to suppress indices when the contractions are clear. The identity contribution is easily calculated via the overlap formulae for coherent states and gives $\< \bm{\phi} (x,t+\epsilon) |\bm{\phi} (x',t)\> =\exp [ -\frac{\epsilon}{2}\sum_{x} (\bm{\phi}^\dagger(x,t)\partial_t \bm{\phi}(x,t) - \partial_t \bm{\phi}(x,t)^\dagger\bm{\phi}(x,t))]$, however the $H_h$ and $H_m$ terms require more attention.

It is simplest to work in momentum space, for which
\begin{equation*}
\hspace{-2cm}H_{h}+H_{m} = \sum_{p,p',j_\mu,k_\mu} \left (\begin{array}{c} \tilde{a}_{j_\mu,p}^\dagger  \tilde{b}_{j_\mu,p}^\dagger \end{array}\right ) \left ( \frac{\tilde{m}^{jk}_0(p-p',t)}{\sqrt{N}} \sigma_z + J^{jk}(t)\delta(p-p') p' \sigma_z^\mu \sigma_x \sigma_z^\mu \right ) \left (\begin{array}{c} \tilde{a}_{k_\mu,p'} \\ \tilde{b}_{k_\mu,p'} \end{array} \right ),
\end{equation*}
where $\tilde{m}^{jk}_0(p,t) = \frac{1}{\sqrt{N}} \sum_x e^{-ipx} m^{jk}_0(x,t)$ and the indices on  $m_0^{jk}$ and $J^{jk}$ are explicitly independent of $\mu$.

Instead of expanding in terms of fermionic/bosonic coherent states of $a_{j_\mu,x}$ and $b_{j_\mu, x}$ we expand in terms of fermionic/bosonic coherent states  of $\tilde{a}_{j_\mu,p}$ and $\tilde{b}_{j_\mu,p}$.  Insertion of the above $H_h+ H_m$ into $\< \bm{\tilde{\phi}}(p,t)| H_h+ H_m|\bm{\tilde{\phi}}(p',t)\> $ gives, to  $O(\epsilon)$, 
  
\begin{eqnarray}
\hspace{-2cm}N^{-1/2}\!\!\sum_{p,p',j_\mu, k_\mu} \tilde{m}^{jk}_0(p-p',t) \big[\tilde{\phi}^*_{j_\mu,a} (p,t)\tilde{\phi}_{k_\mu,a} (p',t)-\tilde{\phi}^*_{j_\mu,b} (p,t)\tilde{\phi}_{k_\mu,b} (p',t))+ \nonumber \\ + \sqrt{N}\delta(p-p')J^{jk} p'  (\tilde{\phi}^*_{j_\mu,a}(p,t) \tilde{\phi}_{k_\mu,b}(p',t) +\tilde{\phi}^*_{j_\mu,b}(p,t) \tilde{\phi}_{k_\mu,a}(p',t) \big] \nonumber \\
\hspace{-2cm}=N^{-1/2}\!\!\sum_{p,p',j_\mu, k_\mu}  
\tilde{m}^{jk}_0(p-p',t) \tilde{\Psi}_{j_\mu}^\dagger(p,t) \sigma_z \tilde{\Psi}_{k_\mu}(p',t)+   \sqrt{N}\delta(p-p')J^{jk}(t)(\tilde{\Psi}_{j_\mu}^\dagger(p,t) p'\sigma_x \tilde{\Psi}_{k_\mu}(p',t)  )  \nonumber \\
\hspace{-2cm}= \sum_{x,j_\mu, k_\mu}  m^{jk}_0(x,t)\Psi_{j_\mu}^\dagger(x,t) \sigma_z \Psi_{k_\mu}(x,t)+   J^{jk}(t)\Psi_{j_\mu}^\dagger(x,t) i\partial_x\sigma_z^\mu\sigma_x\sigma_z^\mu \Psi_{k_\mu}(x,t),
\end{eqnarray}
where $\Psi_{j_\mu}(x,t)=\left(\phi_{j_\mu,a}(x,t), \phi_{j_\mu,b}(x,t) \right)$ is a two-component \textit{auxiliary field} with flavour index $j_\mu$, and where $j_\mu=1, 2, \ldots, D$ for each of the two separate $\mu=0,1$ sectors.

The interaction term can be evaluated in the same way, and we obtain 
$$\<\bm{\phi}(x,t+\epsilon)| \widehat{H}_{\rm int} | \bm{\phi} (x',t) \> = \sum_{x,j_\mu,k_\mu} \left( R^{jk}(x,t) \Psi^\dagger_{j_\mu} (x,t) \Psi_{k_\mu} (x,t)\right)  \widehat{\Psi}_\A^\dagger (x,y=t),$$ 
where the hat on $\widehat{\Psi}_\A^\dagger$ is again to emphasize that the term is an operator on the \emph{physical} system, as opposed to $\Psi_{j_\mu}(x,t)$ which is an auxiliary two component (Grassmann) spinor, with flavor index $j_\mu$.

The total sum in the spatial direction can now be evaluated, and becomes an integral over $x$, which provides us the single time-step amplitude coming from $\widehat{M}_\epsilon (t)$. Once done we can then sum over the time direction to obtain the final expression for the field state. However the process requires care, and in particular a \emph{field rescaling} to preserve finiteness, and so we discuss this in the next section.

\subsection{Field rescaling and the continuum limit}\label{RescalingofField}

For the graph used in the previous section, and also the square-lattice model in the appendix, the two-dimensional contraction across the graph requires a passage to the continuum in two independent directions, and so must be handled carefully. In this section we briefly spell out the technical details showing that we obtain a well-defined two-dimensional action, and we temporarily suppress the flavor-doubling label $\mu$ to reduce our index load.

Recall the basic form of the $2$D PEPS state:
\begin{eqnarray}
|\chi \> &= {\mathcal C} [ A^{r_1}_{(i_1 \cdots k_1 )} \cdots A^{r_M}_{(i_M \cdots k_M )} \cdots ] |r_1 \cdots r_M \cdots \> \nonumber \\
&= \widehat{\mathcal{C} } [\cdots ] |\Omega_\A\>,
\end{eqnarray}
where $(i_M \cdots k_M)$ are a set of contraction indices dependent on the particular graph structure of the state, and $|r_M\>$ is basis state at lattice site $M$. As we have already explained, the contraction can be rewritten as a dynamical process involving the product of transfer operators generating infinitesimal steps
\begin{equation}\label{amp2D}
\widehat{{\mathcal C}} [\cdots] = \<\omega_L|\widehat{M}_\epsilon(l)\widehat{M}_\epsilon(l-\epsilon)\cdots \widehat{M}_\epsilon (\epsilon)\widehat{M}_\epsilon(0)|\omega_R \>.
\end{equation}
For clarity, we restrict to a finite set of points $\{(x,t)\}$ where $x$ runs over $N_x$ points, separated by a distance $\epsilon_x$ and $t$ runs over $N_t$ points, in timesteps of $\epsilon$, and we make explicit all indices.

Once we have introduced coherent state resolutions of the identity at each graph point we have at each value of $t$ a total of $4DN_x$ complex-valued functions to integrate over, coming from
\begin{equation}
\frac{1}{\pi ^{8DN_x}} \int \prod_{k,s,x} d^2 \phi_{k,s}(x,t) |\otimes_{k,s,x} \phi_{k,s}(x,t) \>\<\otimes_{k,s,x} \phi_{k,s} (x,t)| = \I.
\end{equation}

Inserting $N_t+1$ such resolutions into (\ref{amp2D}) gives
\begin{eqnarray}\label{explicitC}
\hspace{-2cm}\widehat{{\mathcal C}}\hspace{0.05cm} [\cdots] = \int d\mu \left [ \prod_{t=0}^{l+\epsilon} \<\otimes_{k_{t+\epsilon},s_{t+\epsilon},x_{t+\epsilon}} \phi_{k_{t+\epsilon},s_{t+\epsilon}} (x_{t+\epsilon}, t+\epsilon) |\hat{M}(t)|\otimes_{k_t,s_t,x_t}\phi_{k_t,s_t} (x_t, t)\>\right ]   \nonumber \\
 \times \<\omega_L|\phi_{k_{l+\epsilon},s_{l+\epsilon}} (x_{l+\epsilon},l+\epsilon)\> \< \phi_{k_0,s_0} (x_0,0)|\omega_R\>,
\end{eqnarray}
with the measure for the integral given by
\begin{equation}
d \mu =\frac{1}{\pi^{8DN_x(N_t+1)}} \prod_{t=0}^{l+\epsilon} \left [ \prod_{k_t,s_t,x_t} d^2 \phi_{k_t,s_t} (x_t, t) \right ].
\end{equation}
 The amplitudes in the integrand of (\ref{explicitC}) have been calculated in the low-energy sector and in this regime we obtain the expression
\begin{equation*}
\hspace{-3cm}\int d\mu \, \, e^{ \sum_{x,t} \epsilon [-\frac{1}{2}(\phi_{k,s}^*(x,t)\partial_t \phi_{k,s} (x,t) - \phi_{k,s}(x,t)\partial_t \phi^*_{k,s} (x,t))+\Psi^\dagger_j(x,t) (J^{jk} \sigma_z^\mu \sigma_x\sigma_z^\mu i\partial_x + m^{jk}_0\sigma_z)\Psi_k(x,t) + R^{jk} \Psi_j^\dagger(x,t) \Psi_k (x,t) \otimes \sys^\dagger (x,y=t)]}, 
\end{equation*}
for the tensor contraction $\widehat{\C} [ \cdots]$, where the matrices $m_0$ and $R$ are allowed to vary smoothly in both $x$ and $t$.

We now take the spatial $\epsilon_x\rightarrow 0$ limit followed by the temporal $\epsilon \rightarrow 0$ limit. However, to keep things well-behaved, we must first rescale the integration variables
\begin{equation}
\phi_{k_\mu,s}(x,t) \rightarrow  \frac{1}{\sqrt{\epsilon_x}} \phi_{k_\mu,s} (x,t).
\end{equation}
Indeed, this rescaling was to be expected since in order to respect the correct commutation/anti-commutation relations: in 1+1 dimensions we have $\Psi_{k_\mu} \sim a_{k_\mu} /\sqrt{\epsilon_x}$, while in 2+1 dimensions we should instead have $\Psi_{k_\mu} \sim a_{k_\mu} /\sqrt{\epsilon_x \epsilon_y}$, where $\epsilon_x$ and $\epsilon_y$ are the two spatial lattice scales. 

Once this rescaling is performed, we find in the $\epsilon_x, \epsilon \rightarrow 0$ continuum limit that the resultant cMPS field state becomes
\begin{equation*}
\hspace{-3cm}|\chi\> = \!\!\int \!\! \D \Psi_{k_\mu} \! \D \Psi^\dagger_{k_\mu} e^{ \int dxdt (-\Psi^\dagger_{k_\mu}(x,t)\partial_t \Psi _{k_\mu}(x,t) +\Psi_{j_\mu}^\dagger(x,t) (J^{jk} \sigma_z^\mu \sigma_x\sigma_z^\mu i\partial_x + m^{jk}_0\sigma_z)\Psi_{k_\mu}(x,t) + (R^{jk}\Psi_{j_\mu}^\dagger(x,t) \Psi_{k_\mu} (x,t))  \sys^\dagger (x,y=t))} |\Omega_\A\> ,
\end{equation*}
and we obtain a well-defined two-dimensional action. 
 
\subsection{The two-dimensional variational class of field states}
Once the rescaling has been performed we may directly integrate by parts, to obtain the class of two-dimensional field states
\begin{equation}\label{general2DcMPS}
\hspace{-2.5cm}|\chi (J, m_0, R)\> \!=\! \int \D \Psi_{k_\mu} \D \Psi^\dagger_{k_\mu} e^{ \int dxdt (\Psi^\dagger_{j_\mu}(x,t)(-\delta^{jk}\partial_t  + J^{jk}(t) \sigma_z^\mu\sigma_x\sigma_z^\mu i\partial_x + m^{jk}_0(x,t)\sigma_z)\Psi_{k_\mu}(x,t) )} |\Phi\>,
\end{equation}
where we have introduced the coherent field state $|\Phi(x,y)\> = \exp [\int dxdt ( \Phi(x,t) \sys^\dagger(x,y=t) -\Phi(x,t) \sys (x,y=t))]|\Omega_\A \>$ with $\Phi(x,t) =R^{jk}(x,t)\Psi_{j_\mu}^\dagger (x,t)\Psi_{k_\mu} (x,t)$, and there is an implicit sum over $\mu$, $j$ and $k$.  

We have obtained this state through the continuum limit of a PEPS contraction, and so (\ref{general2DcMPS}) describes a class of field states that inherits the desirable properties of PEPS, such as entropy/area laws. The  general action given above is not necessarily invariant under $SO(1,1)$. However, Lorentz symmetry for the field can be achieved for a subclass of states in which $J^{jk}(t) =  i\delta^{jk}$. This allows the momentum terms to respect the desired symmetry and we obtain
\begin{equation}\label{DiracAction}
\hspace{-2cm}|\chi (m,R)\> = \int \D \Psi_{k_\mu} \D \bar{\Psi}_{k_\mu} e^{ i\int dxdt (\bar{\Psi}_{j_\mu}( i\delta^{jk}\slashed{\partial} -m^{jk}(x,t))\Psi_{k_\mu} (x,t) } |\Phi(x,y=t) \>,
\end{equation}
where we have also let $m_0^{jk}=-im^{jk}$. For the situation where $m$ is diagonal and constant over the auxiliary spacetime, the coherent field amplitudes are then recognised as Grassmann/complex number path integrals for a set of $2D$ \emph{uncoupled} Lorentz invariant spinor fields. The spinors have the associated gamma matrices $\gamma^0 =\sigma_z$ and $\gamma^1 = i\sigma_y$ obeying the Clifford algebra relations $\{ \gamma^\mu, \gamma^\nu \} = 2 \eta^{\mu \nu} \I$ with $\eta^{\mu \nu} = \rm{diag} (1,-1)$. Also for $\mu=1$ flavor sector we also perform a parity transformation on $\Psi_{k_1}\rightarrow \gamma^0 \Psi_{k_1}$, which is given by $\gamma^0=\sigma_z$ for the 1+1 dimensional case.

Note that, in obtaining this symmetrical state, we were forced to take both $m_0$ and $J$ to be purely imaginary, which means the term $H_h+ H_m$ corresponds to a generator of a \emph{unitary} transformation on the auxiliary system. In particular $U(l,0) = (\I -i\epsilon H(l))( \I -i \epsilon H(l-\epsilon)) \cdots =\T \exp [-i \int_0^l H(t)]$ generates the cMPS state, where $H(t)$ is built out of second-order and fourth-order combinations of creation and annihilation operators for the auxiliary and system fields. Furthermore, the distinctively two-dimensional term is the contribution from $J(t)$, which couples the two spinor degrees of freedom. By setting $J(t) =0 $ we reduce to a diagonal scenario, of the same form as obtained for the one-dimensional cMPS.

Of course $j^\mu = \bar{\psi} \gamma^\mu \psi$ is a conserved current for the free Dirac field, and in particular $\Psi^\dagger \Psi$ is its charge density. Thus, in the case where the auxiliary state has manifest Lorentz symmetry, and where the different flavors decouple, we have the appealing interpretation that the matrix $R$, which is allowed to vary in both $x$ and $t$, couples the densities of the different flavors and dynamically generates the physical field state.

Of course, we could now weaken the conditions and allow more general $J(x,t)$ and $m(x,t)$ to obtain a Dirac action on a 1+1 dimensional spacetime with non-trivial metric. To do so in general would additionally require the modification of the identity term $\I$ to include a well-behaved function $T(x,t)$. Looking back at the analysis, the key feature involved in the derivation is that the auxiliary particles have two degrees of internal freedom and are allowed to hop left or right with some amplitude or remain stationary while flipping an internal (spin) freedom. This is reminiscent of Feynman's `checkerboard derivation' \cite{feynman:1965,schweber:1986} of the Dirac propagator in 1+1 dimensions from a discrete lattice model. There, an electron moves along infinitesimal lightlike trajectories, while jointly flipping direction and spin under a Poisson process with an imaginary rate $1/im$. In light of this, it is not so surprising that we have obtained a Dirac-like action in our continuum limit, although for us a key component is that an expansion in terms of coherent states in the auxiliary time direction either side of an operator $\widehat{M}_\epsilon=\I+\epsilon \widehat{H}_{\rm tot}$ generates a Legendre transformation of a Hamiltonian $\widehat{H}_{\rm tot} [\pi, \phi] $. For a coherent state-expansion, we obtain a Berry-phase term $i \phi^\dagger \partial_t \phi$ from the coherent state overlaps which can be taken as $\pi \partial_t \phi$ where $\pi = i \phi^\dagger$ is the momentum conjugate to $\phi$. 

Another point perhaps worth emphasizing is the physical interpretation of the auxiliary degrees of freedom that emerge in the preceding analysis. This might seem reminiscent of statistical mechanical settings involving interacting lattice models, where the computation of the partition function is often facilitated by the introduction of auxiliary degrees of freedoms that reduce the computation to non-interacting systems.  The partition function can often be computed through some form of approximation, such as a mean-field assumption or method of steepest decent. Such auxiliary fields typically have the interpretation of an effective local external field, or order parameter, which captures the essential local physics of the system. Here, in contrast, the auxiliary degrees of freedom should not be viewed in the same way. We need not specify the physical interactions present in order to define the auxiliary system, but instead the auxiliary system acts as an ``entanglement regulator'', and through the virtual dynamics that sweeps over the whole system, ensures that the physical field state automatically obeys entanglement area laws.

\subsection{Analytic continuation to the Euclidean sector}

We now have a manifestly Lorentz-invariant auxiliary action, however the resultant physical state for $\A$ will have non-trivial entanglement structure in general. As such we would like to analytically continue to the Euclidean sector and arrive at a \emph{rotation invariant} auxiliary action. For second-order actions we can achieve this analytic continuation simply via $t\rightarrow it$, but for spinors subtleties arise. This coordinate transformation, when carried over to the Lorentz transformation, does provide the correct rotation group, but when acting on the spinors themselves results in $\bar{\psi} \psi$ no longer transforming as a scalar. A direct euclideanizing of fermion fields results in a number of problems, such as a loss of hermiticity within the Euclidean propagator. However, these difficulties were overcome by Osterwalder and Schrader \cite{osterwalder:1972a} by making use of a construction that involves fermion doubling where the number of degrees of freedom are doubled so that the spinor and conjugate spinor are independent of each other and transform appropriately under the Lorentz group. However, there exist alternative approaches to euclideanizing the field that do not require this. Instead of analytically continuing the coordinates it is possible to analytically continue the metric itself $\eta^{\mu \nu}  \rightarrow \eta^{\mu \nu} (\theta)$ so that it forms a one-complex-parameter family of metrics interpolating between the Minkowski and the Euclidean one \cite{mehta:1990a}. A more abstract formulation can be achieved by using vielbeins, but for our purpose we work directly with the spacetime metric. 

The appropriate metric is $\eta_{\mu \nu} (\theta) = (\cos 2\theta/|\cos 2 \theta|, -1)$ defined for $0\le \theta \le \pi/2$ except for the singularity at $\theta=\pi/4$, which can be circumvented via extending $\theta$ to be complex. To carry the symmetry group over we also demand that the Clifford Algebra relation $\{ \gamma_\mu (\theta) ,\gamma_\nu (\theta) \} = 2 \eta_{\mu \nu} (\theta) \I$ holds for all $\theta$, and introduce the $\gamma_5(\theta)$ matrix obeying $\gamma_5(\theta)^\dagger = \gamma_5(\theta)$, $\{ \gamma_5 (\theta) , \gamma_\mu (\theta) \} = 0$ and $\gamma_5(\theta)^2 = \I$. A particular parameterized set of gamma-matrices \cite{mehta:1990a} that allow the continuation are then given by
\begin{eqnarray}
	\gamma_1 (\theta) &= \gamma_1 \nonumber \\
	\gamma_0 (\theta) &= \frac{1}{|\cos 2 \theta|^{1/2}} (\gamma_0 \cos \theta + i \gamma_5 \sin \theta ) \nonumber \\
	\gamma_5 (\theta) &= \frac{1}{|\cos 2 \theta|^{1/2}} (\gamma_5 \cos \theta - i \gamma_0 \sin \theta),
\end{eqnarray}
where $\gamma_5 = \sigma_x$ for our 1+1 dimensional case.

This family of gamma matrices obeys the correct anti-commutation relations, and more importantly, generates an interpolation from the $SO(1,1)$ Lorentz transformations to the $SO(2)$ rotations via $\Lambda (\omega; \theta) = \exp[\omega^{\mu \nu}\Sigma_{\mu \nu} (\theta) ]$, where $\omega$ parametrizes the group and $\Sigma_{\mu \nu} (\theta) =\frac{1}{4} [ \gamma_\mu (\theta) , \gamma_\nu (\theta) ]$ are its generators. The intuition behind this choice of parameterization is that to construct scalars under the Lorentz symmetry $SO(1,1)$, fermions in the $(0,\frac{1}{2})$ representation are contracted with ones in $(\frac{1}{2},0)$, whereas for $SO(2)$ we form contractions within the same representation \cite{wukitung}. Since $\gamma_5$ is reducible over the different helicities, while $\gamma_0$ is not, the intuition is to rotate $\gamma_0 \leftrightarrow \gamma_5$ to pass from $SO(1,1)$ to $SO(2)$.

A one-parameter family of actions, symmetric under this group action, can then be constructed, where $\theta=0$ is the Lorentz-invariant Dirac action and $\theta = \pi/2$ is the desired rotation-invariant Euclidean action. It is found to take the form
\begin{equation}
S[\psi, \bar{\psi} ;\theta] = \int d^2x \sqrt{-\det(\eta(\theta))} \psi^\dagger \gamma_0 [ i \eta^{\mu \nu}(\theta) \gamma_\mu (\theta) \partial_\nu - m ] \psi.
\end{equation}

By inspection, we can see that this action could be obtained as the continuum limit of a one-parameter family of discrete tensor networks described by $\hat{M}_\epsilon (t; \theta)$ of the form
\begin{equation}
\widehat{M}_\epsilon(t; \theta) = \I \frac{\cos \theta}{\sqrt{\cos 2 \theta}} - \sigma_y \frac{\sin \theta}{\sqrt{\cos 2\theta}} + \frac{\sqrt{\cos 2 \theta}}{\sqrt{|\cos 2 \theta|}} (H_m + H_{\rm h} + \widehat{H}_{\rm int}).
\end{equation}

The parametrization of the metric has been transferred to the tensor contraction across the discrete graph, and we can smoothly transform from $\theta=0 $ to $\theta= \pi/2$ to obtain, in the continuum limit, the rotation-invariant Euclidean action, and the cMPS class
\begin{equation}\label{eq:euclideancMPS}
|\chi (m,R)\> = \int \D \psi\D \psi^\dagger \exp [-S_E [\psi, \psi^\dagger]] |\Phi\>,
\end{equation}
where $S_E = \int d^2x \psi^\dagger (i \gamma^E_5)(i \slashed{\partial}^E - m) \psi$ is the rotation-invariant action with $\gamma^E_5 = -i \gamma_0$, and the flavor indices are implicit, and are the same as in equation (\ref{DiracAction}).

The above analysis works independent of whether we have used bosonic fields or fermionic fields, however in the latter case it is possible to adopt a related approach that does not require the modification of the gamma matrices. Instead the Grassmann spinor $\psi (x)$ and its conjugate $\bar{\psi}(x)$ are taken independent of each other, and instead of working with the full metric $\eta^{\mu \nu}$ one can analytically continue vielbeins $e^m_\mu \rightarrow e^m_\mu (\theta) = e^{i\theta}\delta^m_\mu $ and use these to construct a Dirac action in the way one would for curved spacetimes \cite{wetterich:2010ni}.

\subsection{Area law properties}

The two-dimensional field state $|\chi\>$ given by (\ref{eq:euclideancMPS}) naturally inherits local properties from the discrete state. In particular, the resultant state is necessarily local in its entanglement structure and obeys an area law. For example, we could consider a finite region $A$ of points on the graph described above. For this region we can define its boundary $\partial A$ as the set of points in $A$ within a distance $\epsilon$ of points not in $A$, and $|\partial A|$ as the number of points in this set. For any pure quantum state $|\psi_{AB}\>$ there exists a unique measure of entanglement, namely the von Neumann entropy of entanglement, or simply the \emph{entanglement entropy}, defined as $S_A = -\Tr[\rho_A \log \rho_A]$ where $\rho_A$ is the reduced state on system $A$.

For discrete states the dimension $D$ of the tensor labels within the contraction places an upper bound on the rank of the reduced state on $A$. The local definition of the state means that only systems in $\partial A$ are entangled with the region $B$ and so the entanglement entropy is upper bounded as $S_A \le c |\partial A|$ where $c$ is a constant dependent on $D$. For the continuum limit we need only impose that the region $A$ is of fixed area with boundary of fixed length $|\partial A|$. Since the number of points in $\partial A$ will scale \emph{linearly} in $1/\epsilon$, we then deduce that $S_A \le c |\partial A| /\epsilon$ and so it is clear that the resultant field state also obeys an area law.

\section{Discussion and Conclusions}

In this paper we have constructed an abstract class of physically natural quantum field states motivated by techniques coming from the discrete regime. Central to the construction is a path-integral representation that we first introduce for one-dimensional cMPS states, before extending to higher dimensions. The class was shown to be complete, in the sense of being able to capture any field state, and efficient in terms of the number of variational parameters. 

The representation allowed us to construct a continuum limit of the PEPS class in two dimensions, and then impose natural, rotational symmetries through the adjustment of the auxiliary action that defines the field state. The desired symmetries of the physical states are encoded in the auxiliary dynamics, and to obtain rotation-invariant states we consider the (imaginary time) evolution of a Lorentz-invariant auxiliary field theory in $(1+1)$ dimensions. To obtain this from first principles we began with a discrete PEPS on a particular graph, and demonstrated that the low-energy sector gave rise to a class of states which manifestly included rotation-invariant states. The continuum limit of the lattice state was shown to produce a doubling of flavors for which the two sectors can be made to simply decouple for PEPS data that varies sufficiently slowly with respect to the scale of the network. Significantly, this doubling only occurs for the auxiliary degrees of freedom while the physical fields are left unaltered. The rotational invariance of the resultant field state arises from the symmetries of an auxiliary spinor action realised as the analytic continuation of a Lorentz-invariant action to imaginary time. Since PEPS states automatically obey area laws, we immediately deduce that so too do the constructed quantum field states.

The one-dimensional cMPS states had been derived in previous work, and so we should ask what is gained by reformulating it as a path integral. Firstly, we gain conceptual insight into the structure and interpretation of the continuum states, in particular we have a natural dynamical description for the virtual/auxiliary degrees of freedom. Secondly, the path integral formulation was useful in the construction of continuum tensor networks beyond one spatial dimension. Finally the formulation of an action for the auxiliary field system has allowed us to impose symmetries in a natural way. Such constructions are much less obvious starting from the alternative representations for 1-d cMPS.

Beyond the utility in constructing higher dimensional field states, and imposing symmetry, we can question whether the path integral formulation benefits the computation of physical expectation values. Here the question is still open, and more work is required on the topic. Generically PEPS are known to be computationally intractable, and so we do not expect that the continuum versions will be any better. However approximate techniques are known to exist in the discrete regime, together with classes of efficiently contractible states. One future method would involve a modification of the general PEPS construction to one that scales in nicer ways. For example, one could imagine building up the two or three dimensional PEPS through local sequential applications of unitaries \cite{Banuls:2008aa,Schoen:aa}. It is known that for such models that the computation of local observables is efficiently contractible. From this the ground-state of an interacting two-dimensional field system with Hamiltonian $H$ could be estimated by first constructing a class of continuum PEPS states that manifest appropriate symmetries, but with each trial wavefunction constructed solely by local sequential unitaries. To determine the ground-state estimate one would then efficiently compute and minimize $\<H\>$ over the variational parameters introduced. The path integral formulation could also benefit from the connection with perturbative methods in field theory. This direction would involve the perturbative expansion of the auxiliary action in terms of its interaction terms, and which would in turn lead to variational parameter regions that increase with the order of perturbation.

Lattice QCD makes use of Monte Carlo sampling of Wilson's Euclidean lattice version of gauge theories,
and has been a remarkably successful method in the computation of the physics of non-perturbative regimes. Despite this, lattice Monte Carlo sampling has downsides. Firstly there is the infamous sign problem that hinders application to states with large fermionic densities, and secondly it faces challenges in describing dynamical scenarios of non-equilibrium systems. Since we have constructed a variational class of field states with no restrictions on the particular statistics of the physical fields, and since variational methods evade the sign problem afflicting Monte Carlo techniques, it would be of interest to see how the cMPS variational class performs in finite fermion density scenarios.  Already there is work \cite{Banuls:2013aa, 2014arxiv1411.0020B, 2015PhRvX...5a1024H} on the extension of discrete MPS states to gauge invariant systems, and therefore it would be of value to connect the present work with these discrete constructions.

\section*{Acknowledgements} 
Helpful discussions with Henri Verschelde are gratefully acknowledged. This work was supported by  EU grants QUERG and QFTCMPS,  FWF SFB grants FoQuS and ViCoM, and by the cluster of excellence EXC 201 Quantum Engineering and Space-Time Research. DJ is supported by the Royal Society.

\newpage
\bibliography{Physics}

\appendix

\section{Continuum limit of the square lattice PEPS}\label{squarelatticeappendix}

It is useful to spell out what exactly happens when we try to proceed directly from the square lattice PEPS to the continuum, and show that the resultant state is not rotationally symmetric, but retains features of the underlying lattice. 

We work with a square lattice of points labeled as $(x,y)$ with the sides of the square of length $\epsilon$. To each vertex there is associated a tensor $A^r_{(ijkl)}$ with $r$ the physical index for the resultant state and the remaining four indices denoting the two directions entering and leaving the point. The indices $i,j,k,l$ take on discrete values $1,\dots D$. 

We shall find that it is convenient to make use of the diagonal coordinates $u= \frac{1}{2} (x-y)$ and $v= \frac{1}{2} (x+y)$ to describe the points in the plane. In these coordinates we have points near $(x,y)$. Note that basic translations in this coordinate system are of the form $(\pm \epsilon/2, \pm \epsilon/2)$. Here $v$ is the auxiliary time direction, while $u$ is the auxiliary spatial direction. 

The basic form of the cMPS state is 
\begin{eqnarray}
|\chi \> = {\mathcal C} [ A^{r_1}_{(i_1j_1 k_1 l_1)} \cdots A^{r_N}_{(i_N j_N k_N l_N)} ] |r_1 \cdots r_N \>
\end{eqnarray}
where ${\mathcal C} [ \cdot ]$ denotes contraction over all indices in accordance with the graph. 

We quantize the indices $ k \in \{1, \dots D\}\mbox{ and } (s=a,b)$, by mapping them to orthonormal states $|k,s\>$, and take these index states as orthogonal one particle modes of a bosonic system $|k,s\> \rightarrow | 1_{k,s} \> := a^\dagger _{k,s} |\Omega_\B \>$. For clarity, we can then view these index states as living in a Hilbert space $\H_a (u,v) \otimes \H_b (u,v)$, associated to the lattice point $(u,v)$. We consider $v=$constant lines, and define
\begin{eqnarray}
\H_{aux} (v) &:=& \otimes_{u \in \mathbb{Z} } \H_a (u,v) \otimes \H_b (u,v)
\end{eqnarray}
where ultimately we will identify $\H_{aux} (v)$ and $\H_{aux} (w)$ for any two times $v$ and $w$. Also, for simplicity we consider a fixed array of indices $B=\{ r (u,v), \cdots, r(p,q), \cdots \} $ defined over the 2D square lattice of points. 
 Graphically, the contraction term is represented as
 
\begin{equation}\label{contractfig}
\hspace{-2cm}	\xymatrix{
	 & & & & & & \\
	 {\mathcal C} [ A^{r_1}_{(i_1j_1 k_1 l_1)} \cdots A^{r_N}_{(i_N j_N k_N l_N)} ]& = &\bullet \ar[dr]_u\ar[ur]^v\ar@{:}[ddddrrrr]&*+[F:<3pt>]\txt{$A$}\ar[d]\ar[l]&*+[F-:<3pt>]\txt{$A$}\ar@{-}[r]\ar@{-}[d]\ar@{-}[l]&*+[o][F-]{\omega_R}\ar@{-}[d]\ar@{-}[l]&\\
&&*+[F:<3pt>]\txt{$A$}\ar[r]\ar[u]&&*+[F-:<3pt>]\txt{$A$}\ar@{-}[r]\ar[l]\ar[d]&*+[F-:<3pt>]\txt{$A$}\ar@{-}[r]\ar@{-}[d]&*+[o][F-]{\omega_R}\ar@{-}[d]\\
&&*+[F:<3pt>]\txt{$A$}\ar@{-}[u]\ar@{-}[d]&*+[F:<3pt>]\txt{$A$}\ar[r]\ar[u]\ar@{-}[l]\ar@{-}[d]&&*+[F:<3pt>]\txt{$A$}\ar[l]\ar[d]&*+[F:<3pt>]\txt{$A$}\ar@{-}[l] \\
&&*+[o][F-]{\omega_L}\ar@{-}[u]&*+[F:<3pt>]\txt{$A$}\ar@{-}[r]\ar@{-}[u]\ar@{-}[l]\ar@{-}[d]&*+[F:<3pt>]\txt{$A$}\ar[u]\ar@{-}[d]\ar[r]&&*+[F:<3pt>]\txt{$A$}\ar[l]\ar@{-}[u]\ar[d] \\
&&&*+[o][F-]{\omega_L}\ar@{-}[r]&*+[F:<3pt>]\txt{$A$}\ar@{-}[r]&*+[F:<3pt>]\txt{$A$}\ar[r]\ar[u]&\\
&&&&&&
}
\end{equation}
where a contraction cut has been made along the $v=$ constant line. Lines with arrows correspond to uncontracted indices, with right/up-pointing ones corresponding to bra-indices and left/down-pointing ones corresponding to ket-indices. The state of the auxiliary system at any fixed timestep is taken along similar diagonal lines, with boundary states $|\omega_L\>$ and $|\omega_R\>$ as shown.

\subsection{The contraction as an inner product}

For any fixed diagonal $v=v_0$ we can split the tensor contraction ${\mathcal C } [\cdots ]$ into two parts and write the result as an inner product $ {\mathcal C} [ \cdots ] = \<\omega^L (v_0) |\omega^R (v_0)\> (B) $

Where  $\< \omega^L (v_0)| $ denotes a `forward pointing' tensor and $|\omega^R (v_0)\>$ denotes a `backward pointing' tensor (see figure \ref{contractfig} ). 
Furthermore, we have that 
\begin{eqnarray}
|\omega^R(v_0) \> \in \H_{aux} (v_0) \nonumber\\
\<\omega^L(v_0) | \in \H^*_{aux} (v_0)
\end{eqnarray}

We capture the dependence of ${\mathcal C} [ \cdots ]$ on $ A^r_{(ijkl)}$ by writing $|\omega^R (v_0) \>= U [\bm{r}(v_0) ] |\omega^R (v_1) \>$. Here we have advanced the cut from $v=v_0$ to $v=v_0 +\epsilon/2$ and have the new `frontier' vector $ |\omega ^R (v_1) \> \in \H_{aux} (v_0+\epsilon /2)$
and a transition operator 
\begin{eqnarray}
U[\bm{r} (v_0)]: \H_{aux} (v_0 +\epsilon /2) \rightarrow \H_{aux} (v_0)
\end{eqnarray}
being a linear operator that describes the advancing contraction cut (or equivalently the infinitesimal evolution of the state $|\omega^R(v_1) \>$), and $\bm{r} (v_0)= (r_1 (v_0), r_2 (v_0), \dots ) $ denotes a list of physical indices along the $v=v_0$ timeslice.

The operator $U[\bm{r} (v_0) ]$ is a string of tensor terms built from an operator $\hat{M}$ (different from the $\hat{M}$ defined in the paper). In particular 
\begin{eqnarray}
U[\bm{r} (v_0) ] &=& [\hat{M} [r_1(v_0)]\otimes \hat{M} [r_2(v_0)]\otimes \hat{M} [r_3(v_0)] \cdots ]
\end{eqnarray}
where crucially, $\hat{M}[r_u (v_0) ]$ is a mapping between the Hilbert spaces
\begin{eqnarray*}
\hspace{-2cm}\hat{M}[r_u(v_0)]: \H_a (u +\epsilon/2, v_0+ \epsilon/2) \otimes \H_b (u-\epsilon/2, v_0+\epsilon/2) \rightarrow \H_a(u,v_0) \otimes \H_b (u,v_0)
\end{eqnarray*}
and is given in terms of the PEPS tensors as
\begin{eqnarray}\label{Mop}
\hat{M}[r_u (v_0) ] &=& \I+  A^{r_u (v_0)}_{(ijkl)} |ij \> \< kl |.
\end{eqnarray}

We may iterate this expansion of $|\omega^R(v_0) \>$ to obtain the expression
\begin{eqnarray*}
{\mathcal C} [\cdots] &=& \<\omega^L(v_0)|U[\bm{r}(v_0)]U[\bm{r}(v_1)]U[\bm{r}(v_2)]U[\bm{r}(v_2)]\cdots U[\bm{r}(v_N)]|\omega^R (v_N) \>
\end{eqnarray*}
where $v_n = v_0+n\epsilon/2$.

\subsection{Construction of the Path Integral}
We assume that at each point $(u,v)$ on the square lattice we have the index state space $\H_a(u,v)\otimes \H_b(u,v)$, and the resolution of the identity operator for this space $\I_{(u,v)}$ given by
\begin{eqnarray}
\I_{(u,v)} &=& \frac{1}{\pi^{2D} }\int \prod_{k,s} d^2 \phi_{k,s} \, \, |\otimes_{k,s} \phi_{k,s} (u,v)\> \< \otimes_{k,s} \phi_{k,s} (u,v)|
\end{eqnarray}
where $|\phi_{k,s} \> = \exp [ \phi_{k,s} \hat{a}^\dagger _{k,s} - \phi^*_{k,s} \hat{a} _{k,s}] |\Omega_\B \> = D(\phi_{k,s}) |\Omega_B \>$.
We also have the identity operator for $\H_{aux} (v)$ as $ \I_{(v)} = \otimes_u \I_{(u,v)}$. Here, we do not consider a single tensor product of Hilbert spaces at every point on the lattice, but instead view things `dynamically' as a tensor product defined along a line $v=$ constant, and so operators like $U(\bm{r}(v))$ and $\I_v$ are linear mappings from $\H_{\rm aux} (v)$ to $\H_{\rm aux} (v+\epsilon/2)$. 
 
To construct the path integral, we begin with
\begin{eqnarray*}
{\mathcal C} [\cdots] &=& \<\omega^L(v_0)|U[\bm{r}(v_0)]U[\bm{r}(v_1)]U[\bm{r}(v_2)]U[\bm{r}(v_2)]\cdots U[\bm{r}(v_N)]|\omega^R (v_N) \>
\end{eqnarray*}
and as before, we insert a resolution of the identity either side of $U[r (v)]$. That is, we consider $\I_{(v+\epsilon/2)} U[\bm{r} (v) ]\I_{(v)}$.
This provides us with amplitudes
\begin{equation*}
\hspace{-2cm}\<\otimes_i \phi_{i,a} (u,v) |\otimes \< \otimes_j \phi_{j,b}(u,v)| M[r_u(v) ] | \otimes_k \phi_{k,a}u+\epsilon/2,v+\epsilon/2) \>  | \otimes_l \phi_{l,b}(u-\epsilon/2,v+\epsilon/2)\>. \nonumber
\end{equation*}

The central amplitude is then 
\begin{eqnarray}
\hspace{-2cm}\<\otimes_k \phi_{k,a}(u,v) |\otimes_k \phi_{k,a}(u+\epsilon/2,v+\epsilon/2) \>\< \otimes_j \phi_{j,b}(u,v)| \otimes_j \phi_{j,b}(u-\epsilon/2,v+\epsilon/2)\>. &\nonumber \\
+  A^{r_u (v)}_{(i k ; j l)} \< \otimes_i \phi_{i,a}(u,v) |1_{i,a} \> \< \otimes_j \phi_{j,b}(u,v)| 1_{j,b} \> \times & \nonumber \\
\times \< 1_{k,a}|\otimes \phi_{k,a} (u+\epsilon/2,v+\epsilon/2) \> \< 1_{l,b} |\otimes \phi_{l,b}(u-\epsilon/2,v+\epsilon/2)\>& \nonumber
\end{eqnarray}

From the formula for the overlap of coherent states, the identity term gives the amplitude
\begin{eqnarray}
 \hspace{-2cm}\exp [-\epsilon/4 (\phi_{k,a} (\partial_u + \partial_v) \phi^*_{k,a}+\phi_{k,b} (-\partial_u + \partial_v) \phi^*_{k,b} - (\phi^*_{k,a} (\partial_u + \partial_v) \phi_{k,a} +\phi^*_{k,b} (-\partial_u + \partial_v) \phi_{k,b}))] \nonumber \\
 \hspace{-1.5cm}+  A^{r_u (v)}_{(i k ; j l)} \< \phi_{i,a}(u,v) |1_{i,a} \> \< \phi_{j,b}(u,v)| 1_{j,b} \> \< 1_{k,a}|\phi_{k,a}(u+\epsilon/2,v+\epsilon/2) \> \< 1_{l,b} |\phi_{l,b}(u-\epsilon/2,v+\epsilon/2)\> \nonumber
\end{eqnarray}
which can be expressed more compactly by defining $\Psi_k := (\phi_{k,a} ,\phi_{k,b})$, to give
\begin{eqnarray}
\hspace{-2cm}\exp [ -\frac{\epsilon}{4} ( (\Psi_k)^\dagger (\sigma_z \partial_u + \I \partial_v ) \Psi^*_k -(\Psi_k)^\dagger (\sigma_z \partial_u + \I \partial_v ) \Psi_k )] + \nonumber \\
 \hspace{-2cm}
+  A^{r_u (v)}_{(i k ; j l)} \< \phi_{i,a}(u,v) |1_{i,a} \> \< \phi_{j,b}(u,v)| 1_{j,b} \> \< 1_{k,a}|\phi_{k,a}(u+\epsilon/2,v+\epsilon/2) \> \< 1_{l,b} |\phi_{l,b}(u-\epsilon/2,v+\epsilon/2)\> \nonumber
\end{eqnarray}

Note that transforming to the original coordinates, $x=u+v$ and $y=v-u$ we then have the first term contributing the amplitude
\begin{eqnarray}
\exp[ i \delta S [ \Psi, \Psi^* ] ]&\equiv &\exp [ -\frac{\epsilon}{4} ( \Psi_k^T \nabla \cdot \Psi^*_k -\Psi^\dagger_k \nabla \cdot \Psi_k )] 
\end{eqnarray}
defining an action $S$ in either the $(u,v)$ coordinates or in the original $(x,y)$ coordinate system. It is clear that this action is not rotationally invariant. This is not particular to PEPS states, but would generically be expected from a continuum construction based on an underlying regular lattice.

The above term is for the identity contribution in $U$. A particular tensor network state on the square lattice is described by $\{A^r_{ijkl} \}$, and so by following the above prescription, we can create the appropriate operators within the one-particle sector of the Fock space.

 A simple and natural candidate is to take $\hat{M}$ of the form $\hat{M} = \mathbb{I} + \frac{\epsilon}{2} (Q_a^{ij}(u,v) \hat{a}^\dagger_{i} \hat{b}_{j} \otimes \mathbb{I}_b + \mathbb{I}_a \otimes Q_b^{ij}(u,v) \hat{b}^\dagger_{i} \hat{b}_{j} )$, where for example $a_{j}, b_k$ are single particle annihilation operators for mode-$j$ in $\H_a$ and mode-$k$ in $\H_b$, and $\{ Q_\alpha^{ij} (u,v) \}_{\alpha, i,j}$ is a set of functions defined over the lattice points. The interpretation of this $Q$-term is that it generates scattering of particles within the internal bond degrees of freedom, without any scattering in the spatial degrees of freedom.

Then for example, if we assume $Q_\alpha^{ij} = Q(u,v)$ we automatically obtain the term 
\begin{eqnarray}
\frac{\epsilon}{2}Q(u,v) \sum_i (|\phi_{a,i}|^2 +|\phi_{b,i}|^2) = \frac{\epsilon}{2}Q(u,v) \Psi^\dagger \Psi
\end{eqnarray}
 in the auxiliary system action after the above expansion in terms of coherent field states.

Of course other options exist for the local contractions. For example, we can consider mixing the spatial modes with a term such as
\begin{eqnarray}
\hat{M}(u,v) &=& \I +\epsilon \hat{R}^{IJ} \hat{a}^\dagger_I \hat{a}_J \nonumber \\
&=& \I + \epsilon \hat{R}^{(i_a,i_b) (j_a j_b)} (\hat{a}^\dagger_{i_a} \otimes \hat{b}^\dagger_{i_b}) (\hat{a}_{j_a} \otimes \hat{b}_{j_b})
\end{eqnarray}
where $\hat{R}^{IJ} (u,v)$ can be taken to be a creation operator for particles in the \emph{physical} field state. Clearly the addition of such a term generates \emph{quartic} terms in the action of $(\phi_a, \phi_b)$.

 The above analysis provides us with an amplitude contribution along a spatial slice, however to then sum over the auxiliary time and obtain the full amplitude requires a re-scaling of the field. The result of this is to yield an action 
\begin{eqnarray}
S = \int dx dy\frac{1}{2} (\Psi^\dagger_k \nabla \cdot \Psi_k-Q(x,y) \Psi^\dagger \Psi ) .
\end{eqnarray}
We have shown the explicit details of this rescaling in section (\ref{RescalingofField}), but there for the case of the rotationally invariant action.

\end{document}